\documentclass[11pt]{JHEP}
\usepackage{latexsym}
\usepackage{amssymb,amsfonts}
\font\mybb=msbm10 at 10pt
\def\bb#1{\hbox{\mybb#1}}
 \preprint{arXiv:1404.1299 [hep-th] April 4, V2: April 28, 2014}
\renewcommand{\theequation}{\arabic{section}.\arabic{equation}}
\title{Twistor/ambitwistor strings and null-superstrings in spacetime of D=4,10 and
11 dimensions}

\author{
Igor Bandos
  \\   Department of
Theoretical Physics, University of the Basque Country
 \\ UPV/EHU,
P.O. Box 644, 48080 Bilbao, Spain
\\ and \\ IKERBASQUE, Basque Foundation for Science, 48011, Bilbao, Spain}

\date{April 4, 2014, V2 April 10, 2014
 printed \today }

\abstract{We show that, at the classical level, the recently proposed `ambitwistor string' model is equivalent to the spinor moving frame formulation of  null-supersting, which in its turn is equivalent to Siegel's formulation of closed twistor string or to its higher dimensional generalizations.
Although the null-(super)string is usually considered as describing the  tensionless limit of (super)string, we show that its action can be derived from the spinor moving frame formulation of superstring also in the infinite tension limit. This observation allows us to argue on the absence of
critical dimensions and suggests that the (ambi)twistor string based technique(s) to calculate field theory amplitudes can be developed  not only in  D=10 or 26,  but also in D=11 and other dimensions.
The D=11 and D=10 twistor strings are described in some details.
}

\begin{document}

\section{Introduction}

In recent  \cite{Mason:2013sva} Mason and Skinner proposed an 'ambitwistor'  string and superstring models and argued that, upon quantization according the prescription discussed in \cite{Mason:2013sva}, these contain only massless particles in the quantum state spectrum,  are consistent in D=10 (D=26 in the bosonic case) and reproduce  the Cachazo-He-Yuan  formulae for  tree-level scattering amplitudes \cite{Cachazo:2013hca}. (In recent \cite{Dolan:2013isa} these formulae for SYM amplitudes were proved using the BCFW (Britto-Cachazo-Feng-Witten) techique \cite{Britto:2005fq}). The NSR version of the ambitwistor string were also discussed in \cite{Mason:2013sva} and in \cite{Adamo:2013tsa}, while the corresponding limit of the pure spinor formulation of superstring was the subject of \cite{Berkovits:2013xba}.

According to \cite{Mason:2013sva} the  ambitwistor string appears as an infinite tension limit of the standard Green-Schwarz (GS) superstring. On the other hand, the authors of \cite{Mason:2013sva} noticed the relation with the equations from the famous papers by Gross and Mende \cite{Gross:1987kza,Gross:1987ar}, which described the string at ultra--high energy and due to this reason, usually associated with tensionless limit of string rather than with the limit of infinite tension.

In this paper we show that, at the classical level, the ambitwistor string model is equivalent to null- superstring as described in  moving frame and spinor moving frame formulation (see \cite{BZ-null,BZ-nulS}\footnote{See \cite{Schild:1976vq,Karlhede:1986wb,Zheltukhin:1997wj,Lindstrom:1990qb,Bozhilov:1999mh} and refs. in \cite{Lindstrom:2003mg} for other formulations of null-string and tensionless string action.} for D=4 case). We also study the  tensionless limit and the limit of infinite tension of the (spinor) moving frame formulation of the  GS superstring, which was proposed in \cite{BZ-str0} and studied  in \cite{BZ-str} \footnote{See \cite{Sok,niss,K+R88,Ghsds,GHT93} other approaches to superparticle and supertring models using the spinor moving frame variables (Lorentz harmonics)}, and  show that, in suitable setups,  both limits  can produce the null-superstring action. This provides us with an explanation why the ambitwistor string, claimed to be the infinite tension limit of the superstring in \cite{Mason:2013sva}, may reproduce the amplitudes with the properties characteristic for  the tensionless limit of the superstring \cite{Gross:1987kza,Gross:1987ar}.

Furthermore, it is known that the tensionless (limit of)  string does not have critical dimensions
(see \cite{Lindstrom:2003mg,Francia:2002pt,Sagnotti:2003qa,Bonelli:2003kh}, refs in \cite{Lindstrom:2003mg} and also \cite{BZ-null,BZ-nulS}). This suggests that the ambitwistor string also can be formulated and is consistent in an arbitrary dimension $D$, including in $D=11$, where its quantization according (a generalization of) the scheme used in \cite{Mason:2013sva,Berkovits:2013xba,Adamo:2013tsa} (or following the line of \cite{Abe:2004ep,Boels:2006ir,Adamo:2013cra}) should produce (tree and one-loop) amplitudes for the 11 dimensional supergravity.

In \cite{Bandos:2006af} it was shown that the D=4 ${\cal N}=4$ version of the null-supestring model \cite{BZ-nulS} is equivalent to the closed twistor string model proposed by Siegel in \cite{Siegel:2004dj} (see \cite{Witten:2003nn} for the original formulation, \cite{Berkovits:2004hg} for another open twistor string action, as well as \cite{Abe:2004ep,Boels:2006ir,Adamo:2013cra,Engelund:2014sqa} and refs. \cite{Adamo:2013cra} in for further development of the twistor string approach).
In this paper we show that spinor moving frame formulation the D-dimensional null--superstring action, which is  classically equivalent to the ambitwistor string,  is also equivalent to the `D-dimensional'  generalization of the Siegel's twistor string action.
Thus ambitwistor string can be also called twistor string.

In contrast with D=4, the  D=10,11 versions of Siegel's twistor string, which we describe in some details,  are formulated in terms of strongly constrained spinors related to the spinor moving frame variables of \cite{BZ-str0,BZ-str,IB+AN=95,IB07:M0}. However, the similarity of null-superstring and superparticle action may simplify the quantization of such a constrained system. An interesting problem for future is to quantize the D=10 and D=11 twistor strings according to the line developed in \cite{IB07:M0} for 11D superparticle, and to compare the results with \cite{Mason:2013sva} and with \cite{Berkovits:2013xba}.

\section{Ambitwistor string action and kappa--symmetry of ambitwistor superstring}
\label{aTwac}

The action for bosonic `ambitwistor string' proposed by Mason  and Skinner in \cite{Mason:2013sva} reads
\begin{eqnarray}\label{SaTw=b}
S^{bosonic}_{MS}=\; \int_{{\cal W}^2}d^2\xi \left( P_a \bar{\partial}X^a-{e\over 2}  P^2\right) \; ,    \qquad
\end{eqnarray}
where ${\cal W}^2$ is the two dimensional worldsheet with local coordinates $\xi^m=(\xi^0,\xi^1)$, $X^a=X^a(\xi)$ is a coordinate function describing the embedding of  ${\cal W}^2$ as a surface in D--dimensional spacetime $M^D$,
$a=0,1,...,(D-1)$, and $P_a=P_a(\xi)$ and $e=e(\xi)$ are auxiliary fields. $\bar{\partial}$ is derivative in one of the directions  of the worldsheet. If the signature is taken to be Euclidean, one can introduce a complex structure and related complex coordinates, $z=\xi^0+i\xi^1$  and $\bar{z}=\xi^0-i\xi^1$, and identify   $\bar{\partial}= {\partial}_{\bar{z}}:= {\partial\over \partial\bar{z}}$. Considering the case of Minkowski signature, one can use $\bar{\partial}= {\partial}_= = {1\over 2}({\partial}_0-{\partial}_1)$. However, in this paper we will often use the notation $\bar{\partial}{}_{\bar{z}}=\bar{\partial}$ also for this case.

Eq. (\ref{SaTw=b}) differs from the massless particle action $S_{0}=\; \int_{W^1}d\tau \left( p_a {\partial}_\tau x^a-{e\over 2} p^2\right)$ only by replacing worldline $W^1$ by the worldsheet ${\cal W}^2$ and allowing all the field depend on two coordinates of the worldsheet.

The {\it ambitwistor superstring} action is obtained by substituting $\bar{z}$ component of the pull--back of the bosonic supervielbein form, $E_{\bar{z}}^a=\bar{\partial}Z^M E_{M}^a(Z)$ for  $\bar{\partial}X^a$ in (\ref{SaTw=}),
\begin{eqnarray}\label{SaTw=}
S_{MS}=\; \int_{{\cal W}^2}d^2\xi \left( P_a E_{\bar{z}}^a -{e\over 2} P^2\right) \;  ,    \qquad
\end{eqnarray}
%\footnote{The NSR-like version of `ambitwistor string' was also proposed in \cite{Mason:2013sva} and studied
% in \cite{Adamo:2013tsa}. We will not discuss it in this paper.}.
In the case of flat target superspace $\Sigma^{(D|n)}$, which we are interested in, the supervielbein can be written in the form
\begin{eqnarray}
\label{Ea=dzEa}
E^a = dX^a - i d\Theta\Gamma^a\Theta
\, , \qquad
E^{\underline{\alpha}}= d\Theta^{\underline{\alpha}}
\;
\end{eqnarray}
where $\Theta^{\underline{\alpha}}$ are the fermionic coordinates of the superspace,
and
\begin{eqnarray}
\label{Ea=dzEa}
E_{\bar{z}}^a = \bar{\partial}X^a - i \bar{\partial}\Theta\Gamma^a\Theta\, , \qquad
E_{\bar{z}}^{\underline{\alpha}}= \bar{\partial}\Theta^{\underline{\alpha}}
\end{eqnarray}
appears in the decomposition of their pull-backs (which we denote by the same symbols) \begin{eqnarray}
\label{Ea=dzEa}
E^a = d\xi^m E_{m}^a =dz E_{{z}}^a+ d{\bar{z}}E_{\bar{z}}^a,\qquad
E^{\underline{\alpha}}= d\Theta^{\underline{\alpha}}= d\xi^m {\partial}_m\Theta^{\underline{\alpha}}=
dz {\partial}\Theta^{\underline{\alpha}}+ d{\bar{z}}\bar{\partial}\Theta^{\underline{\alpha}}\; . \end{eqnarray} These include $D$ bosonic and $n$ fermionic coordinate functions
\begin{eqnarray}
\label{Ea=dzEa}
Z^M(\xi)\equiv Z^M(z,\bar{z})= (X^a(\xi), \Theta^{\underline{\alpha}}(\xi))\; , \qquad
a=0,1,...,(D-1)\; , \qquad \underline{\alpha}=1,..., n \; ,
\end{eqnarray}
where $n$ depends on $D$ and also on ${\cal N}$ in the case of  ${\cal N}$--extended supersymmetry.

For $D=4$, ${\cal N}$-extended superspaces (of which
${\cal N}=4$ case is relevant the maximal 4D SYM theory), $a=0,1,2,3$, $n=4{\cal N}$,
$\Theta^{\underline{\alpha}}=(\theta^{{\alpha}}_i, \bar{\theta}{}^{\dot{\alpha}i})$ with
$\alpha =1,2$, $\dot{\alpha}=1,2$, $i=1,..., {\cal N}$,  and
\begin{eqnarray}
\label{4D4N}
D=4\; :  \qquad \Gamma^a_{\underline{\alpha}\underline{\beta}}= \left(\matrix{0 & \sigma^a_{\alpha\dot{\beta}}\delta^i{}_j \cr
\sigma^a_{\beta\dot{\alpha}}\delta_i{}^j & 0
}\right)
 \; , \qquad with \quad \alpha =1,2\; , \quad \dot{\alpha}=1,2, \quad  i=1,..., {\cal N}\, , \quad
\end{eqnarray}
where $\sigma^a_{\beta\dot{\alpha}}= \epsilon_{\beta\alpha} \epsilon_{\dot{\alpha}\dot{\beta}}
 \tilde{\sigma}{}^{a\dot{\beta}\alpha}$ are relativistic Pauli matrices, $\sigma^a\tilde{\sigma}{}^b+\sigma^b\tilde{\sigma}{}^a= 2\eta^{ab} {\bb I}_{2\times 2}$,  so that
\begin{eqnarray}
\label{Ea=4D}
E^a= dX^a- i d\theta_i\sigma^a\bar{\theta}{}^i  + i d\theta_i\sigma^a\bar{\theta}{}^i
\, . \quad
\end{eqnarray}
In the case of $D=10$, ${\cal N}=1$ superspace, relevant for the 10D SYM theory, $a=0,1,...,9$, ${\underline{\alpha}}=1,...,16$ is the 10D Majorana--Weyl index, the indices of the 16x16 matrix
$\Gamma^a_{\underline{\alpha}\underline{\beta}}=\sigma^a_{\underline{\alpha}\underline{\beta}}=\sigma^a_{\underline{\beta}\underline{\alpha}}$ cannot  be risen, but there exist $\tilde{\sigma}^{a\underline{\alpha}\underline{\beta}}=\tilde{\sigma}^{a\underline{\beta}\underline{\alpha}}= \tilde{\Gamma}^{a\underline{\alpha}\underline{\beta}}$ which obey $\sigma^a\tilde{\sigma}{}^b+\sigma^b\tilde{\sigma}{}^a= 2\eta^{ab} {\bb I}_{16\times 16}$,
\begin{eqnarray}\label{sts+sts=I}
D=10\; : \quad  \Gamma^a_{\underline{\alpha}\underline{\beta}}=\sigma^a_{\underline{\alpha}\underline{\beta}} \; , \quad   \tilde{\Gamma}{}^{a \underline{\alpha}\underline{\beta}}=\tilde{\sigma}{}_a^{\underline{\alpha}\underline{\beta}} \; , \qquad  \underline{\alpha},\underline{\beta}=1,...,16\; ,\qquad
\sigma^a\tilde{\sigma}{}^b+\sigma^b\tilde{\sigma}{}^a= 2\eta^{ab} {\bb I}_{16\times 16} \, . \quad
\end{eqnarray}
Notice also the famous identity
\begin{eqnarray}
\label{sasa=0}
D=10\; :  \qquad \sigma_{a\; \underline{\alpha}(\underline{\beta}}\sigma^a_{\underline{\gamma}\underline{\delta})}\equiv 0
\,  \quad
\end{eqnarray}
which is very important in the Green--Schwarz (GS) superstring model. The classical  GS superstring exists only in the dimensions D=3,4,6, 10 where a counterpart of  (\ref{sasa=0}) is valid. In contrast the existence of this identity is completely irrelevant for the ambitwistor superstring as defined by the action (\ref{SaTw=}).

This is explained by close relation of (\ref{SaTw=}) with the Brink-Schwarz superparticle action $S_{BS}=\; \int_{W^1} \left( P_a E^a -d\tau {e\over 2} P^2\right)$ which basically consists in replacing wirldline $W^1$ by the worldsheet ${\cal W}^2$ and allowing all the fields to depend on two worldsheet coordinates $\xi^m=(\tau, \sigma)$. Indeed, the local fermionic $\kappa$--symmetry, which leaves invariant the action  (\ref{SaTw=}), is similar to the massless superparticle  $\kappa$--symmetry \cite{kappaS}
\begin{eqnarray}
\label{kappaInfR}
 \delta_\kappa \Theta^{\underline{\alpha}}=
P_a \tilde{\Gamma}^{a\underline{\alpha}\underline{\beta}}\kappa_{\underline{\beta}}\; , \qquad \delta_\kappa X^a=i\delta_\kappa \Theta\Gamma^a\Theta\; ,\qquad \delta_\kappa  e=-2i\Theta\kappa\, ,
 \,  \quad
\end{eqnarray}
and relies only on the defining  property of the $\Gamma$--matrix
\begin{eqnarray}\label{sts+sts=I}
(\Gamma^a\tilde{\Gamma}{}^b+\Gamma^b\tilde{\Gamma}{}^a)_{\underline{\alpha}}{}^{\underline{\beta}}= 2\eta^{ab} \delta_{\underline{\alpha}}{}^{\underline{\beta}} \, . \quad
\end{eqnarray}

This implies that, as in the case of massless superparticle \cite{Bergshoeff:1996tu}, the classical `ambitwistor string' of \cite{Mason:2013sva}, as described by the action (\ref{SaTw=}), does exist in target superspace of any bosonic dimension. Below we will discuss an indication that this is true also for the quantum theory.

Thus we can consider also the {\it 11D ambitwistor superstring} characterized by the action (\ref{SaTw=}) with $a=0,1,..., 9,10$,  (\ref{Ea=dzEa}) and
\begin{eqnarray}\label{11D}
D=11\; : \qquad  \Gamma^a_{\underline{\alpha}\underline{\beta}}=\Gamma^a_{\underline{\beta}\underline{\alpha}}
= (C\tilde{\Gamma}^aC)_{\underline{\alpha}\underline{\beta}}\; , \qquad C_{\underline{\alpha}\underline{\beta}}=-C_{\underline{\beta}\underline{\alpha}}
 \; \qquad \; \underline{\alpha},\underline{\beta}=1,...,32\; . \quad
\end{eqnarray}
The quantization of this model along the (generalization of the) line of \cite{Mason:2013sva,Adamo:2013tsa} or  \cite{Berkovits:2013xba} presumably gives the formulae for tree-level amplitudes of 11D supergravity.

This statement is in contradiction with the point of view in \cite{Mason:2013sva,Adamo:2013tsa}, where the stringy  critical dimensions D=26 and D=10 are attributed to the bosonic and supersymmetric versions of ambitwistor string. Below we present some arguments in favor of that the ambitwistor string, being classically equivalent of null--superstring, does not have critical dimensions and can be defined in any D including D=11.

To lighten the notation, from now on we will omit underlining of the Majorana and Majorana-Weyl spinor indices,
\begin{eqnarray}\label{simpl:ual-al}
\underline{\alpha},\underline{\beta},\underline{\gamma}, ... \quad \mapsto \quad {\alpha}, {\beta}, {\gamma}, ... \; , \quad
\end{eqnarray}
in all places where this cannot produce a confusion. We also will tend to use the shorter name {\it ambitwistor string} for the ambitwistor {\it super}string.

\section{Ambitwistor string and null superstring. Moving frame enters the game}
\label{nullS}

\subsection{Moving frame formulation of null superstring as equivalent form of the ambitwistor string action}

The Lagrange multiplier $e$ in (\ref{SaTw=}) produces the mass shell conditions
\begin{eqnarray}\label{p2=0}
P_aP^a=0\; . \quad
\end{eqnarray}
This is an algebraic equation so that its solution, if found, can be substituted into the action; thus one can write the action  (\ref{SaTw=}) as
\begin{eqnarray}\label{SaTw=P2}
S'_{MS}&=&\; \int_{{\cal W}^2}d^2\xi E_{\bar{z}}^a P_a\vert_{ P^2=0}  \nonumber \\
 &=&\; \int_{{\cal W}^2}d^2\xi  P_a\vert_{_{ P^2=0}} (\bar{\partial}X^a + fermions) \;  .   \qquad
\end{eqnarray}
At any point of ${\cal W}^2$ we can chose a suitable  Lorentz frame to  solve
 (\ref{p2=0}) by
\begin{eqnarray}\label{Pa=noncov}
P_{(a)}= \rho (\xi)  \, (1,\; \underbrace{0,\, \ldots \, ,0}_{_{D-2}}\; ,-1)\; .
\end{eqnarray}
The general solution of  (\ref{p2=0}) can be obtained from (\ref{Pa=noncov}) by local (on ${\cal W}^2$) Lorentz rotation. This is to say it has the form
\begin{eqnarray}\label{Pa=Up}
P_a(\xi) = U_a^{(b)}
P_{(b)}=
\rho (\xi) (u_a^{0}- u_a^{(D-1)})=: \rho^{\#} (\xi) u_a^{=}(\xi)\, ,
\end{eqnarray}
where $U_a^{(b)}=(u_a^{0}, u_a^{1}, \ldots  u_a^{(D-1)}$ is a Lorentz group valued matrix,
called the  {\it moving frame matrix}. We prefer to write it in terms of light-like vectors
$ u_a^{\pm\pm}= u_a^0\pm u_a^{D-1}$ and to make notation more compact re-denoting  $u_a^{--}=:u_a^{=}$, $u_a^{++}=:u_a^{\#}$,
\begin{eqnarray}\label{Uab=in}
 U_a^{(b)}(\xi) = \left( {1\over 2}\left( u_a^{=}+u_a^{\#}
 \right), \; u_a^i \, , {1\over 2}\left( u_a^{\#}-u_a^{=}
 \right)\right)\; \in \; SO(1,D-1)\qquad
\end{eqnarray}
This matrix describes a local Lorentz frame adapted to an embedding of the worldsheet ${\cal W}^2$ into the spacetime in such a way that the spacial component of the momentum density at point $\xi$, $P_a(\xi)$, is oriented along the (D-1)-th axis and has a negative projection on this axis.

The fact that the moving frame matrix $U$ belongs to $O(1,D-1)$ is expressed by the
constraint $U^T\eta U=\eta$, which implies that its columns are orthogonal and normalized;  in terms of $u_a^{=}(\xi)=u_a^{0}-u_a^{(D-1)}$ and $u_a^{\#}(\xi) =u_a^{0}+u_a^{(D-1)}$ these properties are expressed by
\begin{eqnarray}\label{uu=0}
&& u_a^{=}u^{a=}=0 \; , \quad \\ \label{uu=2}
 && u_a^{\#}u^{a\#}=0 \; , \quad  u_a^{=}u^{a\#}=2 \; , \quad \\ \label{uui=0} && u_a^{i}u^{a=}=0 =u_a^{i}u^{a\#} \; , \quad  u_a^{i}u^{aj}=-\delta^{ij} \; . \qquad
\end{eqnarray}
Equivalently, the statement of (\ref{Uab=in}), $U_a^{(b)}(\xi) \; \in \; SO(1,D-1)$, can be expressed by  `unity decomposition' $U\eta U^T=\eta$, {\it i.e.}
\begin{eqnarray}\label{I=UU}
\delta_a{}^b= {1\over 2}u_a^{=}u^{b\#}+ {1\over 2}u_a^{\#}u^{b=}-  u_a^{i}u^{bi} \; . \qquad
\end{eqnarray}
The fact that $U$ belongs to $SO(1,D-1)$ subgroup of $O(1,D-1)$ implies that $det U_a^{(b)}=1$ i.e. that
\begin{eqnarray}\label{detU=1}
\epsilon^{abc_1...c_{D-2}}u_b^{=}u_{c_1}^{i_1}...u_{c_{D-2}}^{i_{D-2}}=- (-)^D u^{a=}\epsilon^{i_1...i_{D-2}}\; , \qquad \nonumber \\ \epsilon^{abc_1...c_{D-2}}u_b^{\#}u_{c_1}^{i_1}...u_{c_{D-2}}^{i_{D-2}}=+(-)^D u^{a\#}\epsilon^{i_1...i_{D-2}}\; , \qquad
\end{eqnarray}

The splitting (\ref{Uab=in}) of the moving frame matrix is invariant under local $SO(1,1)\otimes SO(D-2)$ transformations; $u_{a}^{i}$ transforms as $SO(D-2)$ vector,  while the light--like vectors $u_a^=$ and $u_a^\#$ carry the weights $-2$ and $+2$ of  $SO(1,1)$ group,
\begin{eqnarray}\label{SOxSO}
u_{a}^{=} \mapsto u_{a}^= exp{\{ -2\beta\}  }\; , \qquad u_{a}^{\# } \mapsto u_{a}^{\# }  exp{\{2\beta \} }\; , \qquad u_{a}^{i}\mapsto u_{a}^{j}{\cal O}^{ji}\; , \quad {\cal O}{\cal O}^T={\bb I}_{_{(D-2)\times(D-2)}}\; .  \qquad
\end{eqnarray}
To make invariant the solution (\ref{Pa=Up}) of the constraint (\ref{p2=0}), we have to assume that $\rho$ in (\ref{Pa=Up}) carries  $SO(1,1)$ weight $+2$, which is reflected by denoting $\rho=\rho^{\#}$ in the last part of Eq. (\ref{Pa=Up}),
\begin{eqnarray}\label{rhoSO}
\rho^{\# } \mapsto \rho^{\#}  exp{\{2\beta \} }\;  .  \qquad
\end{eqnarray}

Substituting the general solution (\ref{Pa=noncov}) of the constraint (\ref{p2=0}) into (\ref{SaTw=}) (or (\ref{SaTw=P2})) we arrive at following {\it moving frame} action for ambitwistor string
\begin{eqnarray}\label{S=mf}
S_{MF}=\; \int_{{\cal W}^2}d^2\xi  \rho^{\#}  E_{\bar{z}}^{=}:=
\; \int_{{\cal W}^2}d^2\xi  \rho^{\#}  E_{\bar{z}}^a u_a^{=}
 \; .    \qquad
\end{eqnarray}
One can also write this action as an integral of differential form:
\begin{eqnarray}\label{S=mf+}
S_{MF}={i\over 2} \int_{{\cal W}^2}  dz\wedge  \rho^{\#} E^{=} :=
{i\over 2} \int_{{\cal W}^2}  dz\wedge  E_{\bar{z}}^a u_a^{=}  \rho^{\#}
 \; .    \qquad
\end{eqnarray}

Let us compare the above equivalent form of the ambitwistor string action  with (D-dimensional generalization of the) null--superstring action of \cite{BZ-nulS}, which can be written as
\begin{eqnarray}\label{S=BZ}
S_{null}=\; \int_{{\cal W}^2}d^2\xi  \rho^{\# m}  E_{m}^{=}:=
\; \int_{{\cal W}^2}d^2\xi  \rho^{\# m}  E_{m}^a u_a^{=}
 \; .    \qquad
\end{eqnarray}
or, in terms of differential forms, as
\begin{eqnarray}\label{S=BZ+}
S_{null}=\; \int_{{\cal W}^2}\tilde{e}^{\#} \wedge    E^{=}:=
\; \int_{{\cal W}^2}d^2\xi  \tilde{e}^{\#} \wedge  E^a u_a^{=}
 \; ,    \qquad
\end{eqnarray}
where $\tilde{e}^{\#} =d\xi^m e_m^{\#}(\xi )$ and $\tilde{e}_m^{\#}(\xi )\propto \epsilon_{mn}\rho^n$.

The equivalent form (\ref{S=mf}) (and (\ref{S=mf+})) of the  ambitwistor string action  can be considered as a gauge fixed version of the null--superstring action  (\ref{S=BZ})  (and (\ref{S=BZ+})) in which the 2d diffeomorphism gauge symmetry  is broken by the condition
\begin{eqnarray}\label{rhom=rho}
\rho^{\# m}=  \rho^{\#}\delta^m_{\bar{z}}\; ;     \qquad
\end{eqnarray}
the residual part of the diffeomorphism gauge symmetry of the action  (\ref{S=mf}) (and (\ref{S=mf+})) which preserves (\ref{rhom=rho}),  can be identified as  2d conformal symmetry transformations, ${z}\mapsto \bar{f}({z})$, $\bar{\partial} f(z)=0$.

\subsection{Null superstring action from tensionless and infinite tension limits of Green--Schwarz superstring}

\label{aTwS<-SStr}

Now we would like to address the question of what is the limit in which the null--superstring can be obtained from GS superstring. Let us start with (spinor)moving frame formulation of the  GS supersting action \cite{BZ-str0,BZ-str}
\begin{eqnarray}\label{S=GS=mf}
S_{GS}=\; T \int_{{\cal W}^2}\left(
e^{\#}\wedge E^{=} + e^{=}\wedge E^{\#} - e^{\#}\wedge e^{=}\right) - T \int_{{\cal W}^2}{B}_2
 \; .    \qquad
\end{eqnarray}
In it ({\it c.f.} (\ref{S=mf}), (\ref{S=mf+}))
\begin{eqnarray}\label{hE++=}
E^==E^au_a^= \; ,    \qquad E^\# =E^au_a^\# \;     \qquad
\end{eqnarray}
are projections of the pull--back of the target space bosonic supervielbein on the light--like vector fields of the moving frame  (\ref{Uab=in}), $e^\# = d\xi^m e_m^\# (\xi )$ and $e^= = d\xi^m e_m^=(\xi )$ are independent worldvolume 1-forms on  ${\cal W}^2$, $\int_{{\cal W}^2}{B}_2$ is the Wess--Zumino term;  its explicit form, which   is the same  as in the  standard formulation of the GS superstring, will not be needed here. Finally,
\begin{eqnarray}\label{tension=} T={1\over 4\pi \alpha^\prime}\;     \qquad
\end{eqnarray}
is the superstring tension and $\alpha^\prime$ is the Regge slop parameter.
An equivalent form of the (spinor)moving frame action is
\begin{eqnarray}\label{S=GS=mf}
S_{GS}=\; T \int_{{\cal W}^2}d^2\xi \left(
\rho^{\# m} E_m^{=} +  \rho^{= m} E_m^{\#} - \epsilon_{mn} \rho^{\# m} \rho^{= m} \right) - T \int_{{\cal W}^2}{B}_2
 \; ,    \qquad
\end{eqnarray}
where $ \rho^{\# m}\propto \epsilon^{mn} e_n^{\#}$ and $\rho^{= m} \propto \epsilon^{mn} e_n^{=}$.

Clearly, if we set $T=0$ without redefining the basic variables, the action just vanishes.
However, if we first set $e^{\#} =\tilde{e}^{\#}/T= 4\pi\alpha^\prime \tilde{e}^{\#}$ and then take the
$T\mapsto 0$ limit keeping $\tilde{e}^{\#}$ 'fixed', the action  (\ref{S=GS=mf}) reduces to (\ref{S=mf}),
\begin{eqnarray}\label{SGS->Smf0}
\lim\limits_{T\mapsto 0} \left[S_{GS}\vert_{e^{\#}= \tilde{e}^{\#}/T} \right]=
S_{null}
 \; .    \qquad
\end{eqnarray}
This is the reason to consider the null--superstring as tensionless limit of the GS superstring
(and the bosonic null--string as tensionless limit of the bosonic string).

To relate the null superstring, and hence ambitwistor superstring action, to the infinite tension limit of the above action for GS superstring, we need to perform the redefinition with opposite rescaling $e^{\#}= T\tilde{e}^{\#}$ (or $\rho^{\# m}\mapsto  T {\rho}^{\# m}$ ). Then we find
\begin{eqnarray}\label{SGS=T2Smf+}
S_{GS}\vert_{e^{\#}= T\tilde{e}^{\#}} = T^2\left[ S_{null} + {\cal O}({1\over T})\right]\; ,
  \qquad
\end{eqnarray}
which is  dominated by contribution of  null superstring action in the limit of infinite tension,
\begin{eqnarray}\label{SGS:T2->Smf}
\lim\limits_{T\mapsto \infty } \left[ {S_{GS}\vert_{e^{\#}= T\tilde{e}^{\#} } \over T^2} \right] =  S_{null} \; .
  \qquad
\end{eqnarray}

To reproduce the wanted ambitwistor string in this limit we renormalized the action thus making its dimension different from $[\hbar]$($=1$ in our conventions). This is not a problem for classical theory,  as far as in it the action  is used only to derive the equations of motion through the variational principle. In quantum theory this argument does not work as far as the action should be exponentiated. This implies that we need to renormalize the action or our basic variable, $\tilde{e}{}^{\#}$  or  ${\rho}^{\#}$, using another dimensionfull constant.

Notice that such a constant should certainly be present in the ambitwistor string approach of \cite{Mason:2013sva} for calculating scattering amplitudes. Indeed, the proposition of \cite{Mason:2013sva,Adamo:2013tsa} is to calculate the amplitudes for 10D SYM, which is not a conformally invariant theory and has a dimensionfull coupling constant ($[g_{10D}^{YM}]=L^{-3}$)\footnote{This prescription does not work, at least in its literal form, in $D=4$ case where the SYM coupling constant $g_{4D}^{YM}$ is dimensionless, $[g_{4D}^{YM}]=1$. Curiously, this is the case where it had been established that the twistor string produces the tree amplitudes for ${\cal N}=4$ SYM \cite{Witten:2003nn,Berkovits:2004hg}. 
This fact is relevant for our study because, as we show in sec. 5,
the twistor string is equivalent to the ambitwistor string at the classical level. 
On the other hand, studying the loops of the D=4 twistor string, Berkovits and Witten found  \cite{Berkovits:2004jj} that these give rise to the amplitudes of conformal rather than Einstein supergravity. As this is not a consequent theory, the problem of modifying the model in such a way that it's gravity sector is Einstein arose  and was addressed in \cite{AbouZeid:2006wu} (see also \cite{Adamo:2013cra} and refs therein). Such a modification inevitably involves a dimensional constant $\kappa$  of Einstein gravity so that, in the light of  the classical equivalence of twistor string and D=4 null-superstring \cite{Bandos:2006af}, it is tempting to speculate that just  $\kappa$ have to be used to correct the dimension of the variables appearing in the term dominating the infinite tension limit of the superstring action in such a way that it becomes  the dimensionless ($[S]=[\hbar ]$)  null-superstring action. }. It would be interesting to  follow in some details the appearance of the coupling constant in the approach of
\cite{Mason:2013sva,Berkovits:2013xba,Adamo:2013tsa}, but this is certainly out of the score of the present paper\footnote{Let us also stress that ambitwistor string action as proposed in \cite{Mason:2013sva}, Eq. (\ref{SaTw=b}), does possess the conformal invariance which is not the case for the SYM models in $D\not= 4$. This reflects the fact that, besides  the action,  some additional tools are needed to reproduce the correlation functions and scattering amplitudes. This is similar to the situation with recently proposed action \cite{Boulanger:2011dd,Boulanger:2012bj} for the interacting 'higher spin gravity' theories: it  reproduces the 'unfolded' equations by Vasiliev \cite{Vasiliev:1988sa} but, after linearization, does not reproduce the propagators of Fronsdal's  theory \cite{Fronsdal:1978rb}. }. Here we just consider the above discussion as a suggestion that, upon a suitable use of an additional dimensional constant (which is certainly present in the models of interest for the ambitwistor string program)  the infinite tension limit of classical GS superstring can also be described by the null-superstring action.

\section{No critical dimensions for tensionless limit of superstring. Neither for the superstring of infinite tension.  }

It is important that the quantum theory of  tensionless (super)string exists for any spacetime dimensions D \cite{Lindstrom:2003mg,Bonelli:2003kh,Sagnotti:2003qa,BZ-nulS}. In the critical dimensions, where the quantum (super)string with finite tension exists, this can be understood studying the  tensionless limit of quantum superstring, $\alpha^\prime\mapsto \infty$ \cite{Lindstrom:2003mg,Bonelli:2003kh,Sagnotti:2003qa}. To this end one observes that, after a proper normalization of the physical operators of string theory (with dimensionless oscillators $a_n^a, a_{-n\, a}=a^{\dagger}_{n \, a}$ obeying $[a_{n a}, a^\dagger_{m b}]=\delta_{n,m}\eta_{ab}$), the Virasoro generators read (see {\it e.g.} \cite{Bonelli:2003kh}; $a^\dagger_n \cdot a_n :=
a^\dagger_{n b}  a^b_n $)
\begin{eqnarray}\label{L0:=}
L_0&=&{\alpha^\prime\over 2}p_ap^a + \sum_{n>0} na^\dagger_n \cdot a_n \; , \\ \label{Ln=}
L_{n >0}&=& -i \sqrt{2n\alpha^\prime} p_a a_n^a+ \sum_{m>0}  \sqrt{m(m+n)}a^\dagger_m \cdot a_{m+n}
-{1\over 2} \sum\limits_{m=1}^{n-1}  \sqrt{m(m+n)} a_m \cdot a_{n-m}
= (L_{-n})^\dagger\;  .     \qquad
\end{eqnarray}
Notice that, at small tension $T$, this is to say at large $\alpha^\prime$ (\ref{tension=}), the oscillator terms are clearly sub-leading. Then the true symmetry generators in the tensionless limit $\alpha^\prime\mapsto \infty$ are defined by renormalized operators
\begin{eqnarray}\label{LlimT0}L_0/\alpha^\prime &  \longmapsto_{\!\!\!\!\!\!\!\!\!\!\!\!\!\! _{\alpha^\prime\rightarrow  \infty}} & l_0 ={1\over 2}p_ap^a \; , \qquad \\ L_n/\sqrt{\alpha^\prime} & \longmapsto_{\!\!\!\!\!\!\!\!\!\!\!\!\!\! _{\alpha^\prime\rightarrow \infty}}  &  l_n=  -i\sqrt{2n} \,p_a a_n^a\; , \;  .     \qquad
\end{eqnarray}
 which obeys the Heisenberg algebra with central element $l_0$,
\begin{eqnarray}\label{T=0Vir}
[l_n,l_m]=\delta_{n+m}l_0, \qquad [l_{n}, l_0]=0\; , \qquad n\not=0 \; ,      \qquad
\end{eqnarray}
rather than Virasoro algebra\footnote{The critical dimensions are absent also when one quantizes the tensionless string in terms of coordinate functions and conjugate momentum variables  \cite{BZ-null,BZ-nulS}, however in this case the spectrum of mass of the theory is continuous.   }.

Hence tensionless superstring can be defined in spacetime with arbitrary dimension $D$ using string-inspired creation and annihilation operators. With these variables the spacetime conformal invariance is provided by that the quantum state spectrum contains only massless D-dimensional particles, which form  massless supermultiplets in the case of supersymmetric tensionless string.

Now let us discuss  the opposite, infinite tension limit $T={1\over 4\pi \alpha^\prime}\mapsto \infty$ which corresponds to $\alpha^\prime \mapsto 0$. The na\"{\i}ve   limit of the original Virasoro constraints is clearly dominated by  the oscillator terms. But this  contradicts the intuitive expectation that at infinite tension the string looks like particle, so that rather the center of energy motion should be more important.
This suggests to redefine the momentum variable as
\begin{eqnarray}\label{p-tp}
p_a\mapsto  \tilde{p}_a= \alpha^\prime p_a \; , \end{eqnarray} write the Virasoro constraints in this terms, and renormalize them by suitable positive powers of $\alpha^\prime$ before taking the
$\alpha^\prime\mapsto 0$ ($T\mapsto \infty$) limit. In such a way we again reproduce the Heisenberg algebra operators (\ref{T=0Vir})
  \begin{eqnarray}\label{LlimTinf} \alpha^\prime L_0 &\; \longmapsto_{\!\!\!\!\!\!\!\!\!\!\!\!\!\! _{\alpha^\prime\rightarrow  0}} & \quad l_0 ={1\over 2}\tilde{p}_a\tilde{p}^a \; , \qquad \nonumber \\ L_n/\sqrt{\alpha^\prime} & \; \longmapsto_{\!\!\!\!\!\!\!\!\!\!\!\!\!\! _{\alpha^\prime\rightarrow  0}} & \quad l_n=  -i\sqrt{2n} \, \tilde{p}_a a_n^a\; , \qquad \end{eqnarray} but now written in terms of renormalized $\tilde{p}_a$ (\ref{p-tp}).

Certainly the above discussion is not given in terms of variables of the spinor moving frame formulation of null--superstring and has a character of suggestion rather than of the proof; the quantum theory of null-superstring is beyond the scope of this paper.  Notice, however,  that the rescaling (\ref{p-tp}) $p_a\mapsto \tilde{p}_a= \alpha^\prime p_a$  is in correspondence  with
$e^{\#}= T\tilde{e}^{\#}$ (or $\rho^{\#}\mapsto  T{\rho}^{\#}$ ) which was needed to reproduce the $D=3,4,6,10$ null superstring action as a limit of GS superstring action in the corresponding dimension.

The above observations  suggest that null--superstring, which, as we have shown in sec. 3.1, gives one of the classically equivalent formulations  of the ambitwistor string of \cite{Mason:2013sva}, exists in any spacetime dimensions, including say $D=11$, where the tensionfull  superstring (superstring with nonzero tension)  does not exist. Development of a  generalization of  technique of \cite{Mason:2013sva,Berkovits:2013xba,Adamo:2013tsa} for this case might provide us with a tool to calculate tree and one-loop amplitudes of 11D supergravity.

\section{Ambitwistor string as D-dimensional generalization of twistor (super)string. Spinor moving frame enters the game. }
\label{aTwS=TwS}

As it was shown in \cite{Bandos:2006af}, the  $D=4$ ${\cal N}=4$ version of the twistor--like
formulation of null--superstring \cite{BZ-nulS} is equivalent to the closed twistor string
action proposed by Siegel in \cite{Siegel:2004dj} (see \cite{Witten:2003nn} for original formulation and
\cite{Berkovits:2004hg} for an alternative  action for twistor string).
It is natural to expect that  D--dimensional null--superstring, and hence ambitwistor string, also gives rise to a D-dimensional twistor superstring. In this section we show that this is indeed the case for  D=11, D=10 and 4D ${\cal N}$-extended null--superstring.

The spinor moving frame formulation of null--superstrings is related to corresponding  formulation of  massless superparticle in the same manner as originally proposed ambitwistor string action is related to the Brink-Schwarz superparticle action: basically by replacing all the functions of  proper time $\tau$ by the functions of two worldsheet coordinates, $\xi^m=(z,\bar{z})$. Thus the D-dimensional null--superstring actions  can be easily written starting form the spinor moving frame formulations of the massless superparticle actions presented in \cite{B90} for D=4, in \cite{IB+AN=95} for D=3,4,6 and 10 and in
\cite{Bandos:1998wj,Bandos:2006nr,IB07:M0} for D=11.

\subsection{Spinor moving frame }

The moving frame formulation of the null-superstring (\ref{S=BZ})   can be considered as twistor-like spinor moving frame formulation in the following sense.

For every $SO(1,D-1)$ valued matrix $U$, including for the moving frame matrix (\ref{Uab=in}), one can define the matrix $V \in Spin(1,D-1)$ doubly covering $U$ in the sense of
that
 \begin{eqnarray}\label{VGVt=G} V\Gamma_b V^T =  U_b^{(a)} {\Gamma}_{(a)}\, , \quad (a) \qquad V^T \tilde{\Gamma}^{(a)}  V = \tilde{\Gamma}^{b} u_b^{(a)}\;
 \, , \quad (b) \qquad   VCV^T=C \; . \quad (c) \qquad
\end{eqnarray}
These equations reflect the Lorentz invariance of the Dirac (or Pauli) matrices $\Gamma^a$ and of the charge conjugation matrix (when this exists in the minimal spinor representation)

The $SO(1,1)\otimes SO(D-2)$ invariant splitting of the moving frame matrix $U$ is reflected by a splitting of  its doubly covering {\it spinor moving frame matrix} $V$ on two rectangular blocks carrying different $SO(1,1)$ weights and either different or the same representations of $Spin(D-2)$ group. For the 11D case the spinor moving frame matrix reads
\begin{eqnarray}\label{harmVin11}
D=11\; : \qquad V_{(\beta)}^{\;\;\; \alpha}= \left(\matrix{  v^{+\alpha}_q
 \cr  v^{-\alpha}_q } \right) \in Spin(1,10)\;
 \; , \quad \alpha=1,...,32\; , \quad q=1,...,16
 \; . \qquad
\end{eqnarray}
There  16$\times$32 blocks,  $v^{-\alpha}_q$ and  $v^{+\alpha}_q$, called {\it spinor moving frame variables},  carry the opposite $SO(1,1)$ weights and the same $16$ dimensional real (Majorana) spinor representation of $SO(9)$.

In D=10,  the $16\times 16$ spinor moving frame matrix $V$
\begin{eqnarray}\label{harmVin10}
D=10\; : \qquad V_{(\beta)}^{\;\;\; \alpha}= \left(\matrix{  v^{+\alpha}_{\dot{q}}
 \cr  v^{-\alpha}_q } \right) \in Spin(1,9)\; , \qquad \alpha=1,...,16\; , \quad \left\{\matrix{\dot{q}=1,...,8\cr q=1,...,8
 }\right.
 \; . \qquad
\end{eqnarray}
is split on 8$\times$16 blocks,   $v^{+\alpha}_{\dot{q}}$ and $v^{-\alpha}_q$, carrying $c-$ and $s-$ spinorial representation of $SO(8)$.

In both D=11 and D=10 cases, the moving frame variables are strongly constrained by (\ref{VGVt=G}). In particular, the lower diagonal block of (\ref{VGVt=G}a) and $^{(a)}=^=$ component of  (\ref{VGVt=G}b) give the following  constraints involving the light--like vector $u_a^=$,
\begin{eqnarray} \label{Iu--=vGv}
D=10, 11\; : \qquad v_q^{-\alpha} (\Gamma^a)_{\alpha\beta} v_p^{-\beta}= \delta_{qp}
u^{=}_{ a} \; , \qquad   2v_q^{-\alpha} {}v_q^{-\beta} =
u^{=}_{ a}  \tilde{\Gamma}^{a\alpha\beta} \; . \qquad
\end{eqnarray}
Actually, these are constraints for the spinor moving frame variables $v_q^{-\alpha}$, while the vector  $u_a^=$  is defined by these constraints and its property to be light--like is determined by them.
%\left. \qquad \right\} \Rightarrow \quad u^{=a}u_a^{=}=0 \; .
The remaining blocks of the matrix constraint (\ref{VGVt=G}a) and components of (\ref{VGVt=G}b) involve the second spinor moving frame variable $v_{\dot{q}}^{+\alpha}$ (${\dot{q}}$ can be identified with $q$ in $D=11$ case), define the other moving frame vectors, $u^{\#}_a$ and  $u^{i}_a$, and determine their properties listed in (\ref{uu=2}). To resume, (with a suitable gamma matrix representation) the constraints  (\ref{VGVt=G}a,b) imply
\begin{eqnarray}
\label{u--I=vGv}
& v_q^{-} \Gamma^a v_p^{-}= u^{=}_{ a} \delta_{qp}
\; , &\qquad   2v_q^{-\alpha} {}v_q^{-\beta} =
u^{=}_{ a}  \tilde{\Gamma}^{a\alpha\beta} \; ,  \qquad \\
\label{v+v+=u++}
& v_{\dot{q}}^+ {\Gamma}_{ {a}} v_{\dot{p}}^+ = \; u_{ {a}}^{\# } \delta_{\dot{q}\dot{p}}\; , & \qquad 2 v_{\dot{q}}^{+ {\alpha}}v_{\dot{q}}^{+}{}^{ {\beta}}= \tilde{\Gamma}^{ {a} {\alpha} {\beta}} u_{ {a}}^{\# }\; , \qquad
\\ \label{uiG=v+v-}
&
 v_{q}^- {\Gamma}_{ {a}} v_{\dot{p}}^+ = - u_{ {a}}^{i} \gamma^i_{q\dot{p}}\; , &\qquad
 2 v_{q}^{-( {\alpha}}\gamma^i_{q\dot{q}}v_{\dot{q}}^{+}{}^{ {\beta})}=-  \tilde{\Gamma}^{ {a} {\alpha} {\beta}} u_{ {a}}^{i}\; . \quad
\end{eqnarray}
Here, for D=10 $\gamma^i_{q\dot{p}}= \tilde{\gamma}{}^i_{\dot{p}q}$ are the d=8 gamma (actually sigma) matrices, obeying $\gamma^i \tilde{\gamma}{}^j+ \gamma^j \tilde{\gamma}{}^i=\delta^{ij}I_{8\times 8}$ (see e.g. \cite{BZ-str0,BZ-str,IB+AN=95} for their properties), while for D=11  $\dot{q}, \dot{p}$ must  be identified with $q,p$ and the $ 16\times 16$ Dirac matrices $\gamma^i_{q\dot{p}}\equiv \gamma^i_{q{p}}$  are real, symmetric  $\gamma^i_{qp}=\gamma^i_{pq}$ and obey $\gamma^i\gamma^j + \gamma^j \gamma^i= 2\delta^{ij} I_{16\times 16}$ (see \cite{Bandos:2013uoa} for more details).
The constraint (\ref{VGVt=G}c)  allows to express the elements of the inverse 11D spinor moving frame matrix through the same $v_{{q}}^{\pm\beta}$,
\begin{eqnarray}
\label{V-1=CV-A} D=11\; : \qquad  v_{\alpha}{}^{-}_q =  i C_{\alpha\beta}v_{q}^{- \beta }\, ,
\qquad v_{\alpha}{}^{+}_q = - i C_{\alpha\beta}v_{q}^{+ \beta }\, .
 \end{eqnarray}

In D=10 this constraint is  absent (as far as the charge conjugartion matrix does not exists in the D=10 Majorana-Weyl spinor representation) and the elements of the inverse spinor moving frame matrix, $v_{\alpha}^{+q}$ and $v_{\alpha}^{-\dot{q}}$, should be introduced as additional variables
\begin{eqnarray}\label{harmV-1in10}
D=10\; : \qquad V_{\alpha}^{(\beta)}= \left(\matrix{  v_{\alpha}^{+{q}} , & v_{\alpha}^{-\dot{q}}
  }\right) \in Spin(1,9)\; , \qquad \alpha=1,...,16\; , \quad \left\{\matrix{\dot{q}=1,...,8\cr q=1,...,8
 }\right.
 \;  \qquad
\end{eqnarray}
constrained by $V_{\alpha}^{(\beta)}V_{(\beta)}{}^\gamma :=  v_{\alpha}^{-\dot{q}} v^{+\gamma}_{\dot{q}}
+ v^{-\alpha}_q v^{-\gamma}_q
=
\delta_{\alpha}^\gamma$ and
\begin{eqnarray}\label{v-qv+p=}
& v^{-\alpha}_q   v_{\alpha}^{+{p}}=\delta_{qp}\;  , \qquad &
 v^{-\alpha}_q   v_{\alpha}^{-\dot{q}}=0\;  , \qquad \nonumber \\
& v^{+\alpha}_{\dot{q}}  v_{\alpha}^{+{p}}=0\;  , \qquad &
v^{+\alpha}_{\dot{q}} v_{\alpha}^{-\dot{p}} = \delta_{\dot{q}\dot{p}}
 \; .  \qquad
\end{eqnarray}
In terms of the elements of inverse spinor moving frame matrix the constraints (\ref{u--I=vGv})--(\ref{uiG=v+v-}) read
\begin{eqnarray}
\label{v+v+=u++I}
& v_q^{+} \tilde{\Gamma}{}^a v_p^{+}= u^{\#}_{ a} \delta_{qp}
\; , &\qquad   2v^{+q}_{\alpha} {}v^{+q}_{-\beta} =
u^{\# }_{ a}  {\Gamma}^{a}_{\alpha\beta} \; ,  \qquad \\
\label{u--I=vGvI}
& v_{\dot{q}}^- \tilde{\Gamma}_{ {a}} v_{\dot{p}}^- = \; u_{ {a}}^{= } \delta_{\dot{q}\dot{p}}\; , & \qquad 2 v^{-\dot{q}}_{{\alpha}}v^{-\dot{q}}_{ {\beta}}= {\Gamma}^{a}_{ {\alpha} {\beta}} u_{ {a}}^{\# }\; , \qquad
\\ \label{uiG=v+v-I}
&
 v_{q}^+ \tilde{\Gamma}_{ {a}} v_{\dot{p}}^- =  u_{ {a}}^{i} \gamma^i_{q\dot{p}}\; , &\qquad
 2 v^{+q}_{(\alpha |}\gamma^i_{q\dot{q}}v^{-\dot{q}}_{|\beta )}={\Gamma}^{a}_{ {\alpha} {\beta}} u_{ {a}}^{i}\; . \quad
\end{eqnarray}

In the case of $D=4$, $Spin(1,3)=SL(2, {\bb C})$ so that  the spinor moving frame can be defined by  complex  unimodular 2$\times$2 matrix,
\begin{eqnarray}\label{harmVin4}
D=4\; : \qquad V_{(\beta)}^{\;\;\; \alpha}= (V_{(\dot{\beta})}^{\;\;\; \dot{\alpha}} )^*=\left(\matrix{  v^{+\alpha}
 \cr v^{-\alpha}} \right) \in SL(2,{\bb C})\; , \qquad \alpha=1,2\; , \quad \dot{\alpha}=1,2
 \;  \qquad
\end{eqnarray}
so that $v_\alpha^-$ can be considered as nonvanishing, but in all other respect unconstrained complex Weyl spinor, while $v_\alpha^+$ is restricted by the normalization condition $v^{-\alpha}v_{\alpha}^+=1$ only \footnote{In our notation $\epsilon^{12}=-\epsilon_{12}=1$ so that $det V=-\epsilon_{\alpha\beta}v^{+\alpha}v^{-\beta}=v^{-\alpha}v_{\alpha}^+$.},
\begin{eqnarray}\label{v-v+=1}
D=4\; : \qquad && det(V_{(\beta)}^{\;\;\; \alpha})= 1 \qquad \Leftrightarrow \qquad v^{-\alpha}v_{\alpha}^+=1 \nonumber \\ && det(V_{(\dot{\beta})}^{\;\;\; \dot{\alpha}})= 1 \qquad \Leftrightarrow \qquad v^{-\dot\alpha}v_{\dot \alpha}^+=1 \; .
\qquad
\end{eqnarray}
This can be put in correspondence with (\ref{VGVt=G}c), while Eqs. (\ref{VGVt=G}a,b) just define the set of light--like moving frame vectors   as direct products of  $v_{\alpha}^\pm$ and $\bar{v}{}_{\dot{\alpha}}^\pm=(v_{\alpha}^\pm)^*$ \footnote{The set of our light-like moving frame 4--vectors can be recognized as Newmann--Penrose  or isotropic tetrade of \cite{Newman:1961qr,Penrose:1987uia}, $l_a= u_a^{=}$, $n_a=u_a^{\#}$, $m_a=(\bar{m})^a=u_a^{+-}=(u_a^{+-})^*$; the spinor moving frame variables (called Lorentz harmonics in \cite{B90}) can be identified with {\it diads}, $(v_\alpha^{+}, v_\alpha^{-})= ({o}_{\alpha}, {\imath }_\alpha )$ (see \cite{BZ-nulS}).  In (\ref{u++=4D}), to write compact expressions we denoted
$u_a^{\#}=: u_a^{++}$ and $u_a^{=}=:u_a^{--}$. There and below $v^{\pm}\sigma_a\bar{v}^{\pm} = v^{\alpha\pm} \sigma^a_{{\alpha}\dot{\alpha}}\bar{v}{}^{\dot{\alpha}\pm} $, $v^{-}\sigma_a\bar{v}^{+} = v^{\alpha -} \sigma^a_{{\alpha}\dot{\alpha}}\bar{v}{}^{\dot{\alpha}+}$.}
\begin{eqnarray}\label{u++=4D}
D=4\; : \qquad && u_a^{\pm \pm} = v^{\pm}\sigma_a\bar{v}^{\pm} =(u_a^{\pm \pm})^*\; \quad
\Leftrightarrow \quad
u_a^{\pm \pm}\sigma^a_{\alpha\dot{\alpha}} =2 v^{\pm}_{\alpha} \bar{v}{}^{\pm}_{\dot{\alpha}}
 \; , \qquad
\\ && u_a^{-+} = v^{-}\sigma_a\bar{v}^{+} = (u_a^{+-})^* \; \quad \Leftrightarrow \quad
u_a^{\pm \mp}\sigma^a_{\alpha\dot{\alpha}} =2 v^{\pm}_{\alpha} \bar{v}{}^{\mp}_{\dot{\alpha}}
 \; .  \qquad
\end{eqnarray}

\subsection{Spinor moving frame action for null--superstring in D=4 and Siegel's form of the  twistor string action}

We denote the coordinates of ${\cal N}$-extended $D=4$ superspace by $(X^a, \theta^\alpha_i,\bar{\theta}{}^{\dot{\alpha} i})$ with $a=0,1,2,3$, $\alpha=1,2$, $\dot{\alpha}=1,2$ and $i=1,...,{\cal N}$.  The action for the spinor moving frame formulation of  null-superstring  in this superspace can be obtained from the corresponding moving frame action, which in its gauge fixed version has the form   (\ref{S=mf}), (\ref{S=mf+}), by using $u_a^{=} = v^{-}\sigma_a\bar{v}^{-}$,
\begin{eqnarray}\label{S=smf4D}
S_{smf}^{4D}=\;
\; \int_{{\cal W}^2}d^2\xi  \rho^{\#}  E_{\bar{z}}^a u_a^{=}= \int_{{\cal W}^2}d^2\xi  \rho^{\#}  v_{\alpha}^{ -} \bar{v}{}^-_{\dot{\alpha}} \left(\bar{\partial}X^a\tilde{\sigma}{}_a^{\dot{\alpha}{\alpha}}-2i \bar{\partial}\theta_i^\alpha \bar{\theta}{}^{\dot{\alpha} i}+ 2i \theta_i^\alpha \bar{\partial}\bar{\theta}{}^{\dot{\alpha} i}\right)
 \; .    \qquad
\end{eqnarray}
It is similar to the superparticle action considered in \cite{Shirafuji:1983zd} which is known to allow for change of variables to Ferber supertwistor   \cite{Ferber:1977qx}
\begin{eqnarray}\label{cZ=}
 {\cal Z}_{\Lambda} = (\Upsilon_{\underline{\alpha}}; \eta_i)=  ( \lambda_\alpha , \mu^{\dot{\alpha}} ;   \eta_i )\; ,
\end{eqnarray}
and its conjugate
\begin{eqnarray}\label{bcZ=}
\bar{{\cal Z}}{}^{\Lambda} := \Xi^{\Lambda\Pi^* }({\cal Z}_{\Pi})^\dagger = \left(\matrix{\bar{\Upsilon}{}^{\underline{\alpha}}\cr i\bar{\eta}^i}\right)\; =  \left(\matrix{\bar{\mu}{}^\alpha \cr  -\bar{\lambda}_{\dot{\alpha}} \cr i \bar{\eta}^i}\right)\;
\end{eqnarray}
subject to the so--called (super--)helicity constraint
\begin{eqnarray}\label{constr4D}
 {\cal Z}_{\Lambda} \bar{{\cal Z}}{}^{\Lambda}:= \Upsilon_{\underline{\alpha}} \bar{\Upsilon}{}^{\underline{\alpha}} +i \eta_i  \bar{\eta}^i=  \lambda_\alpha \bar{\mu}{}^\alpha - \mu^{\dot{\alpha}}  \bar{\lambda}_{\dot{\alpha}} +i \eta_i  \bar{\eta}^i= 0
 \; .    \qquad
\end{eqnarray}
The relation between supertwistors and the variables of the spinor moving frame action (\ref{S=smf4D})
is given by the Ferber--Penrose incidence relations \cite{Ferber:1977qx}
\begin{eqnarray}\label{l=Xmu}
 \mu^{\dot{\alpha}}=  \lambda_\alpha (X^{{\alpha}\dot{\alpha}} + 2i\theta_i^{{\alpha}} \sigma^a \bar{\theta}{}^{\dot{\alpha}i}) =(X^a + i\theta_i \sigma^a \bar{\theta}^i) \tilde{\sigma}{}_a^{\dot{\alpha}{\alpha}} \lambda_\alpha \; , \qquad \eta_i= 2 \theta^\alpha_i\lambda_\alpha \; , \qquad
\end{eqnarray}
supplemented by
\begin{eqnarray}\label{l=rv}
  \lambda_\alpha = \sqrt{\rho^{\#}} v_\alpha^- \; , \qquad  \bar{\lambda}_{\dot\alpha} = \sqrt{\rho^{\#}} \bar{v}_{\dot\alpha}^{\; -} \; . \qquad
\end{eqnarray}
Notice that (\ref{l=Xmu}) provides the general solution of (\ref{constr4D}).

Now it is easy to check  (see e.g. \cite{Bandos:2006af}) that the  action (\ref{S=smf4D}) is equivalent to
\begin{eqnarray}\label{S=4DTw}
S_{smf}^{4D}=
-  \int_{{\cal W}^2}d^2\xi  (  \bar{\partial}\lambda_\alpha \; \bar{\mu}{}^\alpha -  \bar{\partial} \mu^{\dot{\alpha}}  \; \bar{\lambda}_{\dot{\alpha}} +i  \bar{\partial} \eta_i \;  \bar{\eta}^i )= -
\int_{{\cal W}^2}d^2\xi   \bar{\partial} {\cal Z}_{\Lambda} \;  \bar{{\cal Z}}{}^{\Lambda} \; .
 \;     \qquad
\end{eqnarray}
with the supertwistor variables obeying (\ref{constr4D}).  On the other hand, Eq. (\ref{S=4DTw}) coincides with Siegel's proposal for the twistor (super)string action \cite{Siegel:2004dj}. Thus \cite{Bandos:2006af}   twistor string can be identified with 4D null--superstring  and also, in the present perspective, with the D=4 version of the ambitwistor string model of  \cite{Mason:2013sva}.

The special interest to the ${\cal N}=4$ version of the twistor string model was related to the fact that the superspace spanned by ${\cal N}=4$  supertwistors  allows for the existence of the holomorphic integral measure \begin{eqnarray}\label{4Dmeasure}
 \Omega_{(3|4)} =\epsilon_{\underline{\alpha}\underline{\beta}\underline{\gamma}\underline{\delta}} \Upsilon^{\underline{\alpha}}d\Upsilon^{\underline{\beta}} \wedge d\Upsilon^{\underline{\gamma}}\wedge d\Upsilon^{\underline{\delta}}  \epsilon^{ijkl}{\partial\over \partial \eta^i }{\partial\over \partial \eta^j }{\partial\over \partial \eta^k }{\partial\over \partial \eta^l }\;     \qquad
\end{eqnarray}
invariant under the natural $U(1)$ gauge transformations acting on supertwistors. However, in the perspective of ambitwistor string proposal of \cite{Mason:2013sva} neither this, nor the spacetime superconformal symmetry of the D=4 supertwistor action does play central role, as it has been oriented on description of amplitudes of 10D SYM, which is not conformally invariant.

However, the classical equivalence of the $D=4$ ${\cal N}=4$ ambitwistor string to the twistor string which is known to be consistent \cite{Witten:2003nn,Berkovits:2004hg} (at least in tree approximation) suggests  that this is the case also for the former, and thus gives us one more argument in favor of the conjecture that ambitwistor string exists and is consistent in a spacetime of any dimensions.

Below we describe the D=10 twistor string model which is classically equivalent to the 10D supersymmetric ambitwistor string  of \cite{Mason:2013sva}, and also D=11 twistor string model which describes the 11D generalization of the ambitwistor string. The arguments of the previous sections allow us to expect that this latter should be  consistent and, upon application of the quantization method similar to the one(s) used in  \cite{Mason:2013sva,Berkovits:2013xba,Adamo:2013tsa}, should generate the tree and one--loop amplitudes of 11D supergravity. Actually, we prefer to begin with the D=11 model.

\subsection{Twistor string in D=11 and D=10 from ambitwistor/null superstring action}

To make manifest the twistor-like structure of the moving fame action (\ref{S=mf}) of  D=11 or D=10 null superstring, which, as we have shown, is classically equivalent to ambitwistor superstring, we have to substitute the
expression for $u_a^=$ in terms of spinor bilinear which follow from the constraints
(\ref{u--I=vGv}) or  (\ref{u--I=vGvI}).

\subsubsection{Spinor moving frame action for 11D ambitwistor superstring and its reformulation in the  generalized 11D superspace}

In D=11 it is convenient, following \cite{Bandos:2006nr},  to introduce
\begin{eqnarray}\label{11Dl=rv}
\lambda_{\alpha q}=\sqrt{\rho^{\#}}v_{\alpha}^{-q} \; ,
 \;     \qquad
\end{eqnarray}
which, as a consequence of (\ref{u--I=vGv}) obey \footnote{Remember that in 11D $ v_{\alpha}{}^{-}_q =  i C_{\alpha\beta}v_{q}^{- \beta }$,
$v_{\alpha}{}^{+}_q = - i C_{\alpha\beta}v_{q}^{+ \beta }$, Eq.(\ref{V-1=CV-A}).}
\begin{eqnarray}
\label{lGl=pI}
& \lambda_q \tilde{\Gamma}_{ {a}} \lambda_p = \; P_{ {a}} \delta_{{q}{p}}\; , & \qquad 2 \lambda_{{\alpha}q}\lambda_{ {\beta}q}= {\Gamma}^{a}_{ {\alpha} {\beta}} P_{ {a}}\; , \qquad P_aP^a=0\; . \qquad
\end{eqnarray}
Here the light-like vector $P_a$ originates in the product of the Lagrange multiplier and the moving frame vector,  $P_a=\rho^{\#} u_a^=$, {\it cf.} Eq. (\ref{Pa=Up}).

The relation with spinor moving frame, (\ref{11Dl=rv}), guarantees the consistency of the constraints (\ref{lGl=pI}) (which probably  is not so apparent).

Notice also that all the constraints (\ref{lGl=pI}), as well as all the expressions below,  are invariant under $SO(16)$ transformations acting on the indices $q,p=1,...,16$, so that we can solve the constraints by
\begin{eqnarray}\label{11Dl=rvS}
\lambda_{\alpha q}=\sqrt{\rho^{\#}}v_{\alpha}^{-p}S_{pq} \; , \qquad  S_{pp'}S_{qp'}=\delta_{pq}
 \;   .  \qquad
\end{eqnarray}

Now substituting ${1\over 32}\lambda_{{\alpha}q}\lambda_{{\beta}q}\tilde{\Gamma}^{a {\alpha} {\beta}}$ for
$\rho^{\#}u_a^=$ in (\ref{S=mf}) we obtain
\begin{eqnarray}\label{S=mf=llD11}
S^{11D}_{smf}&=& \int_{{\cal W}^2}d^2\xi  \lambda_{{\alpha}q}\lambda_{{\beta}q} {1\over 32}\tilde{\Gamma}_a^{{\alpha} {\beta}}  \left(\bar{\partial}X^a- i \bar{\partial} \Theta\Gamma^a \Theta  \right)
 = \nonumber  \\
&=& \int_{{\cal W}^2}d^2\xi  \lambda_{{\alpha}q}\lambda_{{\beta}q} \left( \bar{\partial} X^{{\alpha} {\beta}} - i \bar{\partial} \Theta^{(\alpha}\; \Theta^{\beta )}\right)
 \; .    \qquad
\end{eqnarray}
Here to pass to the second form of the action we have defined
\begin{eqnarray}\label{Xab=XGa}
 X^{{\alpha} {\beta}} = {1\over 32} X^a \tilde{\Gamma}_a^{{\alpha} {\beta}}
 \;     \qquad
\end{eqnarray}
and have used the Fiertz identity \begin{eqnarray}\label{11DFiertz}
{1\over 32}\tilde{\Gamma}_a^{{\alpha} {\beta}}   \bar{\partial} \Theta\Gamma^a \Theta  =
\bar{\partial} \Theta^{(\alpha}\; \Theta^{\beta )} + {1\over 64}\tilde{\Gamma}_{ab}^{{\alpha} {\beta}}   \bar{\partial} \Theta\Gamma^{ab} \Theta - {1\over 32\cdot 5!}\tilde{\Gamma}_{abcde}^{{\alpha} {\beta}}   \bar{\partial} \Theta\Gamma^{abcde} \Theta
 \;     \qquad
\end{eqnarray}
to simplify the second term in the bracket. Notice that, as a consequence of (\ref{lGl=pI}),  after contraction with $\lambda_{{\alpha}q}\lambda_{{\beta}q}$ the second and the third terms in (\ref{11DFiertz}) do not contribute so that the net result is provided by the first term, $\bar{\partial} \Theta^{(\alpha}\; \Theta^{\beta )}$.

Furthermore, due to the same reasons one can consider  $X^{\alpha\beta}$ in (\ref{S=mf=llD11}) as generic symmetric matrix. Indeed, the general decomposition of such a $32\times 32$ symmetric matrix
\begin{eqnarray}\label{Xab=XG}
 X^{{\alpha} {\beta}} = {1\over 32} X^a \tilde{\Gamma}_a^{{\alpha} {\beta}} - {1\over 64}\tilde{\Gamma}_{ab}^{{\alpha} {\beta}} Z^{ab} + {1\over 32\cdot 5!}\tilde{\Gamma}_{abcde}^{{\alpha} {\beta}}  Z^{abcde}
 \;     \qquad
\end{eqnarray}
contains the contributions of ''tensorial central charge coordinates''  $Z^{ab}=- Z^{ba}=Z^{[ab]}$ and
$Z^{abcde} =Z^{[abcde]}$. However, these do not produce  any contribution into the action (\ref{S=mf=llD11}) because, as a consequence of (\ref{lGl=pI}),
\begin{eqnarray}\label{lG2l=0}
\lambda_q\tilde{\Gamma}_{ab}\lambda_q =0\; , \qquad \lambda_q\tilde{\Gamma}_{abcde}\lambda_q =0
\; .      \qquad
\end{eqnarray}
(cf. with the discussion of 11D superparticle action in \cite{Bandos:2006nr}).

Hence, interestingly enough, the twistor-like spinor moving frame formulation of ambitwistor/null superstring allows to treat it as a dynamical system in the enlarged superspace of 528 bosonic and 32 fermionic dimensions, $\Sigma^{(528|32)}$ with coordinates $X^{\alpha\beta}$ and $\Theta^\alpha$.
This enlarged superspace was discussed for the first time in \cite{vanHolten:1982mx}. Various  dynamical system in this superspace were studied in \cite{Curtright:1987zc,Bandos:1998vz,Bandos:1998wj,Chryssomalakos:1999xd,Bandos:2003ng}. Notice that it is also related with hidden gauge symmetry structure of 11D supergravity
\cite{D'Auria:1982nx,Bandos:2004xw,Bandos:2004ym} and with notion of BPS preon \cite{Bandos:2001pu,Bandos:2003us}; this allows us to hope on its possible significance in the ambitwistor/twistor string context.

\subsubsection{11D twistor string action from ambitwistor superstring}

Moving the derivative $\bar{\partial}$ we can write the 11D null/ambitwistor   superstring action (\ref{S=mf=llD11}) in the form  characteristic for Siegel's twistor string,
\begin{eqnarray}\label{S=11DtwS}
S_{twS}^{11D}&=& - \int_{{\cal W}^2} \left(
\bar{\partial} \lambda_{{\alpha}q}\;\mu^{\alpha}_q -
\lambda_{{\alpha}q}\bar{\partial}  \mu^{\alpha}_q + i \bar{\partial} \eta_q \; \eta_q
 \right) = \qquad \\ \label{S=11DtwZ}
 &=&- \int_{{\cal W}^2} \bar{\partial}{\cal Z}_{\Lambda q} \Xi^{\Lambda\Sigma} {\cal Z}_{\Sigma q}
 \;   ,  \qquad \Xi^{\Lambda\Sigma} =\left(\matrix{ 0 & \delta^{\alpha}{}_{\beta} & 0 \cr
                                       - \delta_{\alpha}{}^{\beta} & 0 & 0  \cr
                                                    0 & 0 & i } \right)
\end{eqnarray}
The dynamical variables of this action, real bosonic spinors
$\lambda_{{\alpha}q}$ and $ \mu^{\alpha}_q$, and real  fermionic scalars $\eta_q$
can be collected in 16 strongly constrained  $OSp(1|64)$ supertwistors
\footnote{Generically such  supertwistors with 64 bosonic and 1 fermionic components provide the fundamental representation of the $OSp(1|64)$ supergroup. The $OSp(1|64)$ transformations leave invariant the matrix $\Xi^{\Lambda\Sigma}$ defined in (\ref{S=11DtwZ}). See \cite{Bandos:1998vz,Bandos:2003ng} for more details.  }
\begin{eqnarray}\label{11Dstw}
{\cal Z}_{\Lambda q} := (\Upsilon_{\underline{\alpha} q}; \eta_q) =(\lambda_{{\alpha}q}\; , \mu^{\alpha}_q   \; ; \eta_q)
\;  ,   \qquad \alpha=1,...,32\; , \quad q=1,...,16\; .
\end{eqnarray}
They are expressed through the coordinate functions and spinor moving frame variables by the following generalization of the Penrose incidence relations
\begin{eqnarray}\label{mu=lX}
\mu^{\alpha}_q= X^{\alpha\beta} \lambda_{{\beta}q} - {i\over 2} \Theta^{\alpha}\Theta^{\beta}\lambda_{\beta q}
\; , \qquad \eta_q=  \Theta^{\beta}\lambda_{\beta q}\; . \qquad
\end{eqnarray}
These relations {\it with generic $X^{\alpha\beta}=X^{\beta\alpha}$} (\ref{Xab=XG}) and $\lambda_{\alpha q}$ obeying (\ref{lGl=pI}),
\begin{eqnarray}
\label{lGl=pID11}
& \lambda_q \tilde{\Gamma}_{ {a}} \lambda_p = \; P_{ {a}} \delta_{{q}{p}}\; , & \qquad 2 \lambda_{{\alpha}q}\lambda_{ {\beta}q}= {\Gamma}^{a}_{ {\alpha} {\beta}} P_{ {a}}\; , \qquad P_a:=
{1\over 16} \lambda_q \tilde{\Gamma}_{ {a}} \lambda_q\; ,
\end{eqnarray}
provide us with the general solution of the constraints (see \cite{Bandos:2006nr})
\begin{eqnarray}\label{Jc=0}
{\bb J}_{pq}= {\cal Z}_{\Lambda [p} \Xi^{\Lambda\Sigma} {\cal Z}_{q]\Sigma }= 2\lambda_{{\alpha}[p}\;\mu^{\alpha}_{q]} + i  \eta_p \; \eta_q=0 \; . \qquad
\end{eqnarray}
Expression (\ref{mu=lX}) with particular $X^{\alpha\beta}$ expressed through the standard 11D spacetime coordinate, (\ref{Xab=XGa}), appears if we impose, in addition to (\ref{Jc=0})
(\ref{lGl=pID11}), also the constraint
\begin{eqnarray}\label{K=0}
{\bb K}_{pq}= \lambda_{{\alpha}(p}\;\mu^{\alpha}_{q)}- {1\over 16} \delta_{pq} \lambda_{{\alpha}p'}\;\mu^{\alpha}_{p'} =0 \; .   \qquad
\end{eqnarray}
Clearly, this breaks  the $OSp(1|64)$ invariance which might be attributed to  the action (\ref{S=11DtwS}), (\ref{S=11DtwZ}) and to the constraint (\ref{Jc=0}).
To write  (\ref{K=0}) in terms of the whole supertwistor (\ref{11Dstw}) we have to use the degenerate  symmetric matrix
\begin{eqnarray}\label{GLS=}
  G^{\Lambda\Sigma} =\left(\matrix{ 0 & \delta^{\alpha}{}_{\beta} & 0 \cr
                                        \delta_{\alpha}{}^{\beta} & 0 & 0  \cr
                                                    0 & 0 & 0 } \right)
\; ;    \qquad  {\bb K}_{pq}=  {\cal Z}_{\Lambda [p} G^{\Lambda\Sigma} {\cal Z}_{q]\Sigma }= 0\; .
\end{eqnarray}
This  matrix is preserved by  bosonic $GL(32)$ subgroup of $OSp(1|64)$ supergroup only.
But even this invariance is actually broken already by the constraints  (\ref{lGl=pID11}) imposed on the first,  $\lambda$--component  of the supertwistor.

Notice that
\begin{eqnarray}\label{I+=}
{\cal I}_+^{\Lambda\Sigma} &:=&
  {1\over 2}\left( G^{\Lambda\Sigma} + \Sigma^{\Lambda\Sigma} \right)=\left(\matrix{ 0 & 0 & 0 \cr
                                        \delta_{\alpha}{}^{\beta} & 0 & 0  \cr
                                                    0 & 0 & i } \right) \qquad \\ \label{I-=} && or \qquad
                                                    {\cal I}_-^{\Lambda\Sigma}:=
 {1\over 2}\left( G^{\Lambda\Sigma} - \Sigma^{\Lambda\Sigma} \right) =
 \left(\matrix{ 0 & \delta^{\alpha}{}_{\beta} & 0 \cr
             0 & 0 & 0  \cr
                                                    0 & 0 & -i } \right)
  \qquad
\end{eqnarray}
can be considered as a counterpart of D=4 'infinite twistor' ${\cal I}_+^{\underline{\alpha}\underline{\beta}}=
\left(\matrix{ 0 & 0  \cr
                                         0 & \epsilon^{\dot{\alpha}\dot{\beta}} } \right)
$ or  ${\cal I}_-^{\underline{\alpha}\underline{\beta}}=
\left(\matrix{ \epsilon^{{\alpha}{\beta}}   & 0  \cr
                                         0 & 0}\right)
$ \cite{Penrose:1987uia} (see also e.g. \cite{AbouZeid:2006wu}).

Although our supertwistors are strongly constrained, their relation with spinor moving frame variables, Eq. (\ref{l=rv}), allows to define variational problem in a simple way. The most essential is the variation of $\lambda_{\alpha q}$ which can be written in the form of (see \cite{IB07:M0,Bandos:2013uoa} for details)
\begin{eqnarray}\label{vlaq=}
 \delta \lambda_{\alpha q}= i_\delta f^{(0)}  \lambda_{\alpha q} + \lambda_{\alpha p}i_\delta f^{[pq]} +
 {1\over 2} i_\delta f^{- i}  v_{\alpha p}{}^+ \gamma^i_{pq'} S_{q'q}
 \qquad
\end{eqnarray}
where $ i_\delta f^{(0)}$,  $i_\delta f^{[pq]}=-i_\delta f^{[qp]}$ and $i_\delta f^{-i}$ ($= \sqrt{\rho^{\#}}i_\delta \Omega^{= i}$ in the notation of \cite{IB07:M0,Bandos:2013uoa}) denote independent variations and the last term refers explicitly on the solution (\ref{11Dl=rvS})  of the constraints (\ref{lGl=pI}): in it  $S_{pq}=S^{-1}_{qp}$ is the $SO(16)$--valued matrix field, $ v_{\alpha p}{}^+$ are the spinor moving frame variables complementary to  $v_{\alpha p}{}^-$ (see (\ref{harmVin11}), (\ref{v+v+=u++I}), (\ref{u--I=vGvI}), (\ref{uiG=v+v-I})) and  $\gamma^i_{pq}=\gamma^i_{qp}$ are SO(9) invariant gamma matrices.

The twistor-like formulation of 11D massless superparticle model, which can be obtained from our  twistor superstring by replacing worldsheet ${\cal W}^2$ by a worldline, making all the field dependent on one proper time variable $\tau$ instead of two worldsheet coordinates, and replacing  $\bar{\partial}$ by ${\partial}_\tau$, was discussed  in \cite{Bandos:2006nr}, where it was shown that its quantization gives the linearized 11D supergravity multiplet. This allows us to hope  that the quantization of the spinor moving frame formulation of the ambitwistor string model along the lines of \cite{Mason:2013sva,Berkovits:2013xba} and/or \cite{Abe:2004ep,Boels:2006ir,Adamo:2013cra} can produce the amplitudes of the 11D supergravity.

\subsubsection{10D twistor string action from ambitwistor superstring}

The 10D version of the ambitwistor superstring action can also be rewritten in twistor-like (spinor moving frame) form and as a twistor string action for a set of strongly constrained twistors\footnote{
See \cite{Uvarov}  for the discussion on twistor transform of tensionful superstrings in D=3,4,6,10 and \cite{Fedoruk} for related studies.}.
The equations are very similar to the ones for 11D ambitwistor string, up to that the spinor indices $\alpha, \beta$ are now taking $16$ values (not $32$ as in D=11), and the basic set of bosonic spinor carries  the dotted, c-spinor index of $SO(8)$, $\dot{q}=1,...,8$. This allows us to be short in this section.

Using the D=10 spinor moving frame variables we introduce the set of 8 composed Majorana-Weyl spinors,
 \begin{eqnarray}\label{10Dl=rvS}
\lambda_{\alpha \dot{q}}=\sqrt{\rho^{\#}}v_{\alpha \dot{p}}^{\; \; -}S_{\dot{p}\dot{q}} \; , \qquad  SS^T={\bb I}_{8\times 8}
 \;   ,   \qquad \alpha, \beta =1,...,16
\end{eqnarray}
which  solves the constraints ({\it cf.}  (\ref{lGl=pI}))
\begin{eqnarray}
\label{lGl=pID10}
& \lambda_{\dot{q}} \tilde{\sigma}_{ {a}}\lambda_{\dot{p}} = \; p_{ {a}} \delta_{\dot{q}\dot{p}}\; , & \qquad 2 \lambda_{{\alpha}\dot{q}}\lambda_{ {\beta}\dot{q}}= {\Gamma}^{a}_{ {\alpha} {\beta}} p_{ {a}}\; , \qquad
\end{eqnarray}
involving a ten vector field $p_a$.  This is light--like, $p_ap^a=0$, as  can be deduced from the identity $\tilde{\sigma}^{a\alpha (\beta} \tilde{\sigma}^{\gamma\delta )}_{ {a}}=0$ (in D=11, there is no such type identity, and the proof of light-likeness of $P_a$ in (\ref{lGl=pID10}) requires to use the charge conjugation matrix, see {\it e.g.} \cite{Bandos:2013uoa}).

The spinor moving frame action for 10D twistor string reads
 \begin{eqnarray}\label{S=mf=llD10}
S^{10D}_{smf}&=& \int_{{\cal W}^2}d^2\xi  \lambda_{{\alpha}\dot{q}}\lambda_{{\beta}\dot{q}} {1\over 16}\tilde{\sigma}_a^{{\alpha} {\beta}}  \left(\bar{\partial}X^a- i \bar{\partial} \Theta\sigma^a \Theta  \right)
 = \nonumber  \\
&=& \int_{{\cal W}^2}d^2\xi  \lambda_{{\alpha}\dot{q}}\lambda_{{\beta}\dot{q}} \left( \bar{\partial} X^{{\alpha} {\beta}} - i \bar{\partial} \Theta^{(\alpha}\; \Theta^{\beta )}\right)
 \; ,    \qquad
\end{eqnarray}
where in the second line $16\times 16$ matrix field  $ X^{{\alpha} {\beta}}=  X^{ {\beta}{\alpha}}$  can be considered as constructed  from the 10-vector coordinate function,
\begin{eqnarray}\label{Xab=Xsa}
 X^{{\alpha} {\beta}} = {1\over 16} X^a \tilde{\sigma}_a^{{\alpha} {\beta}}
 \; ,     \qquad
\end{eqnarray}
or to be a generic symmetric spin-tensor field
\begin{eqnarray}\label{Xab=XG10}
 X^{{\alpha} {\beta}} = {1\over 16} X^a \tilde{\sigma}_a^{{\alpha} {\beta}}  + {1\over 32\cdot 5!}\tilde{\sigma}_{abcde}^{{\alpha} {\beta}}  Z^{abcde}
 \;     \qquad
\end{eqnarray}
which  contains the  contributions of  5-rank self--dual tensorial
$Z^{abcde}={1\over 5!}\epsilon^{abcdea'b'c'd'e'f'} Z_{a'b'c'd'e'f'}$. This latter does not contribute to the action because the bosonic spinor $\lambda_{\dot{q}}$ in (\ref{10Dl=rvS}) obeys
\begin{eqnarray}\label{ls5l=0}
\lambda_{\dot{q}}\tilde{\sigma}_{abcde}\lambda_{\dot{q}} =0\;   .  \qquad
\end{eqnarray}

As in eleven-dimensional case, the spinor moving frame action of 10D ambitwistor/null string (\ref{S=mf=llD10}) can be equivalently written as twistor string action
\begin{eqnarray}\label{S=10DtwS}
S^{twS}_{10D}&=& - \int_{{\cal W}^2} \left(
\bar{\partial} \lambda_{{\alpha}\dot{q}}\;\mu^{\alpha \dot{q}} -
\lambda_{{\alpha}\dot{q}} \bar{\partial}  \mu^{\alpha \dot{q}} + i \bar{\partial} \eta_{\dot{q}} \; \eta_{\dot{q}}
 \right) = - \int_{{\cal W}^2} \bar{\partial}{\cal Z}_{\Lambda {\dot{q}}} \Xi^{\Lambda\Sigma} {\cal Z}_{\Sigma {\dot{q}}}
 \;   ,  \qquad
\end{eqnarray}
where $\Xi^{\Lambda\Sigma}$ has the same form as in (\ref{S=11DtwZ}), but with $16\times 16$ blocks $\delta_\alpha^\beta$.

The action (\ref{S=10DtwS}) is written in terms of a set of 8 strongly constrained  $OSp(1|32)$ supertwistors
\begin{eqnarray}\label{10Dstw}
{\cal Z}_{\Lambda {\dot{q}}} := (\Upsilon_{\underline{\alpha} {\dot{q}}}; \eta_{{\dot{q}}}) =(\lambda_{{\alpha}{\dot{q}}}\; , \mu^{\alpha}_{\dot{q}}   \; ; \eta_{\dot{q}})
\;  ,   \qquad \alpha=1,...,16\; , \quad {\dot{q}}=1,...,8\;
\end{eqnarray}
which can be expressed through   the 10D coordinate functions and spinor moving frame variables by the following generalization of the Penrose incidence relation
\begin{eqnarray}\label{mu=lX10D}
\mu^{\alpha\dot{q}}= X^{\alpha\beta} \lambda_{{\beta}\dot{q}} - {i\over 2} \Theta^{\alpha}\Theta^{\beta}\lambda_{\beta {\dot{q}}}
\; , \qquad \eta_q=  \Theta^{\beta}\lambda_{\beta {\dot{q}}}\; . \qquad
\end{eqnarray}
In addition to (\ref{lGl=pID10}), the supertwistor components obey the constraint
\begin{eqnarray}\label{Jc=0D10}
{\bb J}_{\dot{p}\dot{q}}= {\cal Z}_{\Lambda [\dot{p}} \Xi^{\Lambda\Sigma} {\cal Z}_{\dot{q}]\Sigma }= 2\lambda_{{\alpha}[\dot{p}}\;\mu^{\alpha}_{\dot{q}]} + i  \eta_{\dot{p}} \; \eta_{\dot{q}}=0 \; , \qquad
\end{eqnarray}
and, in the case of $
 X^{{\alpha} {\beta}}$ expressed through $X^a$ by (\ref{Xab=XG10}), also
\begin{eqnarray}\label{K=0D10}
{\bb K}_{\dot{p}\dot{q}}= \lambda_{{\alpha}(\dot{p}}\;\mu^{\alpha}_{\dot{q})}- {1\over 16} \delta_{\dot{p}\dot{q}} \lambda_{{\alpha}\dot{p}'}\;\mu^{\alpha}_{\dot{p}'} =0 \; .   \qquad
\end{eqnarray}

These relations {\it with generic $X^{\alpha\beta}=X^{\beta\alpha}$} (\ref{Xab=XG}) and $\lambda_{\alpha \dot{q}}$ obeying the constraints
(\ref{lGl=pID10})
provide us with the general solution of the constraints (see \cite{Bandos:2006nr})
\begin{eqnarray}\label{Jc=0D10}
{\bb J}_{\dot{p}\dot{q}}= {\cal Z}_{\Lambda [\dot{p}} \Xi^{\Lambda\Sigma} {\cal Z}_{\dot{q}]\Sigma }= 2\lambda_{{\alpha}[\dot{p}}\;\mu^{\alpha}_{\dot{q}]} + i  \eta_{\dot{p}} \; \eta_{\dot{q}}=0 \; . \qquad
\end{eqnarray}
Eq. (\ref{mu=lX}) with particular $X^{\alpha\beta}$ expressed through the standard 11D spacetime coordinate, (\ref{Xab=XGa}), appears if we impose, in addition to (\ref{Jc=0})
(\ref{lGl=pID11}), also the constraints
\begin{eqnarray}\label{K=0D10}
{\bb K}_{\dot{p}\dot{q}}= \lambda_{{\alpha}(\dot{p}}\;\mu^{\alpha}_{\dot{q})}- {1\over 16} \; \lambda_{{\alpha}\dot{r}}\;\mu^{\alpha}_{\dot{r}} \; \delta_{\dot{p}\dot{q}} =0 \; .   \qquad
\end{eqnarray}

The twistor string  is classically equivalent to the 10D ambitwistor superstring of \cite{Mason:2013sva}, (\ref{SaTw=}) [as its action  (\ref{S=10DtwS}) was obtained from (\ref{SaTw=}) by solving the algebraic constraints and  changing the variables]. We can show that the quantization of its superparticle limit gives the 10D SYM theory. As a result, we can expect that its quantization in the twistor string line of \cite{Abe:2004ep} or in the line of \cite{Mason:2013sva} can   produce the amplitudes of the 10D SYM.

Such a quantization is a natural continuation of our study. An apparent  technical problem on this way is the constrained nature of our supertwistors. However, it can be resolved using their relation with the spinor moving frame  variables: this allows to define the admissible variation of the $\lambda$--spinor  which preserve the constraints
(\ref{lGl=pID10}) which can be written as ({\it cf.} with (\ref{vlaq=}) of the  D=11 case)
\begin{eqnarray}\label{vladq=}
 \delta \lambda_{\alpha \dot{q}}= i_\delta f^{(0)}  \lambda_{\alpha \dot{q}} + \lambda_{\alpha \dot{p}}i_\delta f^{[\dot{p}\dot{q}]} +
 {1\over 2} i_\delta f^{- i}  v_{\alpha p}{}^+ \gamma^i_{p\dot{s}} S_{\dot{s}\dot{q}}\; .
 \qquad
\end{eqnarray}
Here $ i_\delta f^{(0)}$,  $i_\delta f^{[pq]}=-i_\delta f^{[qp]}$ and $i_\delta f^{-i}$  denote independent variations and the last term refers explicitly to the solution (\ref{10Dl=rvS})  of the constraints (\ref{lGl=pI}):  $S_{\dot{p}\dot{q}}=S^{-1}_{\dot{q}\dot{p}}$ is the matrix field taking values in c-spinor representation of  $SO(8)$,  $ v_{\alpha p}{}^+$ are the spinor moving frame variables complementary to  $v_{\alpha \dot p}{}^-$ (see (\ref{harmVin10}), (\ref{v+v+=u++I}), (\ref{u--I=vGvI}), (\ref{uiG=v+v-I})) and  $\gamma^i_{p\dot{q}}=\tilde{\gamma}{}^i_{\dot{q}p}$ are SO(8) invariant gamma matrices.

\section{Conclusion}

In this paper we first have shown that D=4,10 and D=11 versions of the recently proposed (for D=10)   ambitwistor superstring  \cite{Mason:2013sva}\footnote{The D=26 purely bosonic ambitwistor string and the D=10 NSR version of ambitwistor string were also considered in \cite{Mason:2013sva}. } is classically equivalent to the null--superstring in its moving frame and spinor moving frame formulations \cite{BZ-nulS}.

The null--string and null--superstring is usually associated to tensionless limit of Nambu-Goto (NG) string and Green--Schwarz (GS)  superstring. In contrast, the ambitwistor (super)string was associated in \cite{Mason:2013sva} with infinite tension limit of the  NG (GS super)string. However, on the other hand, the authors of \cite{Mason:2013sva} noticed the similarity of the  properties of the ambitwistor string amplitudes with the very high energy limit of string amplitudes  \cite{Gross:1987kza,Gross:1987ar}, which is naturally associated to the tensionless limit of superstring.

To clarify this issue in this paper we have shown how the null--superstring action can dominate both the tensionless and infinite tension limit of the superstring action. Furthermore, the tensionless limit of superstring theory does not have critical dimensions, this is to say, it can be formulated in any D \cite{Lindstrom:1990qb,Lindstrom:2003mg,Francia:2002pt,Sagnotti:2003qa,Bonelli:2003kh}. In this paper we have presented arguments which  suggest that this is also true in the limit of infinite tension. (Actually this conjecture looks quite natural as  in the infinite tension limit string is expected to behave like a particle).

This conjecture is further supported by the  observation that the spinor moving frame formulation of  D=4, ${\cal N}$--extended supersymmetric null--superstring, which is classically equivalent to  D=4 version of the  ambitwistor string, is also classically equivalent to  Siegel's formulation of closed twistor superstring \cite{Siegel:2004dj}. The twistor string was originally formulated for  $D=4$ ${\cal N}=4$ case, in which the twistor superspace is Calabi-Yau  supermanifold \cite{Witten:2003nn,Berkovits:2004hg}, and is known to be a consistent theory (at least at the tree level, see \cite{Berkovits:2004jj,AbouZeid:2006wu} and refs. therin).

The similar twistor transform allows us to present  the spinor moving frame formulation of D-dimensional null superstring, classically equivalent to D-dimensional version of the ambitwistor string, as a D-dimensional twistor string. Besides $D=4$, we have described the D=10 twistor string, classically equivalent to the original 10D ambitwistor superstring action of \cite{Mason:2013sva}, and the D=11 twistor superstring. Both  D=10 and D=11 models are formulated in terms of highly constrained $OSp(1|2n)$ supertwistors ($n=16$ for $D=10$ and $n=32$ for $D=11$). However, the relation of the component of supertwistors  with spinor moving frame variables allows us to present simple expressions for their admissible variations which preserve the constraints.
The quantization  of the 11D null-superstring/twistor string model in the line of \cite{Mason:2013sva} or \cite{Abe:2004ep} is expected to produce amplitudes of 11D supergravity. To develop such a quantization and to obtain the 11D amplitudes  is an interesting task for future study.

Interestingly enough, the generalized Ferber--Penrose (FP) incidence relations expressing the supertwistors through  the coordinate functions and spinor moving frame variables describing the ambitwistor string (null--superstring) are gauge equivalent to a more general FP relations involving additional coordinates of enlarged or tensorial  superspace, $\Sigma^{(528|32)}$ parametrized by $X^a$, $Z^{ab}=Z^{[ab]}$ and $Z^{abcde}=Z^{[abcde]}$ in 11D case. The gauge symmetry of the twistor like formulation of the ambitwistor string and of the  twistor superstring action, which allow to gauge away the additional 517(=55+462)  coordinates $Z^{ab}$ and $Z^{abcde}=Z^{[abcde]}$ seems to be related with the hidden gauge symmetry structure of 11D supergravity \cite{D'Auria:1982nx,Bandos:2004xw,Bandos:2004ym}. This makes tempting to speculate on  possible relevance of the enlarged 'tensorial' superspaces in the ambitwistor string context.

In the recent paper \cite{Bjerrum-Bohr:2014qwa} the authors, approaching  the scattering equation of \cite{Cachazo:2013hca} in the context of (standard) string theory the authors formulated  a new `dual resonance' model which coincides with string theory in both the $\alpha^\prime \mapsto 0$ and $\alpha^\prime \mapsto \infty$ limits. They observed that the  solutions of this dual model can be found algebraically on the surface of scattering equations producing the $\alpha^\prime$ corrections to the amplitudes,  checked that these coincide with known results in several previously studied cases. The authors of \cite{Bjerrum-Bohr:2014qwa} have stress that, despite they worked according the rules of string theory, their dual model cannot by the usual string theory.

In the light of our present results, an interesting question for future study is whether the dual model of
\cite{Bjerrum-Bohr:2014qwa} can be related to (10D verison of) twistor string which, as we have shown, is classically equivalent to null superstring  and ambitwistor string.

{\it Notice added:} After this paper appeared on the net, the ambitwistor string technique have been applied to the D=4 case in \cite{Geyer:2014fka} and used there to obtain the expressions for ${\cal N}$-extended SYM and Einstein supergravity amplitudes. The authors of \cite{Geyer:2014fka} reformulated the ambitwistor string model in terms of supertwistors, similarly to what had been done in secs. 3.1, 5.1, 5.2 above, and noticed that their theory coincides with (``is superficially identical to'') twistor string, thus agreeing with  our conclusion in sec. 5.2. On the other side they  have stressed that the resulting formulae for amplitudes are different from that following from twistor string approach as formulated in \cite{Witten:2003nn,Berkovits:2004hg,Berkovits:2004jj}.  The study of \cite{Geyer:2014fka} confirms the conclusion of this paper on the existence of generalizations of ambitwistor string technique \cite{Mason:2013sva,Adamo:2013tsa} for arbitrary spacetime dimensions.

{\bf Aknowledgments}.
The author thanks Dima Sorokin for numerous useful discussions and communications, and also for hospitality in Padova University and Padova section of INFN, where a part of this work was done. It is also a pleasure to thank Nicolya Boulanger for discussions on \cite{Boulanger:2011dd,Boulanger:2012bj} and \cite{Adamo:2013cra}, which renewed his interest to the twistor string approach. A partial support from the Spanish MINECO research grants FPA2012-35043-C02-01, from the Basque Government Research Group Grant ITT559-10 and from UPV/EHU under the program UFI 11/55 is greatly acknowledged.


\bigskip


\begin{thebibliography}{99}
\renewcommand{\theequation}{R.\arabic{equation}}

%\cite{Mason:2013sva}
\bibitem{Mason:2013sva}
  L.~Mason and D.~Skinner,
  ``Ambitwistor strings and the scattering equations,''
  arXiv:1311.2564 [hep-th].
  %%CITATION = ARXIV:1311.2564;%%


%\cite{Cachazo:2013hca}
\bibitem{Cachazo:2013hca}
  F.~Cachazo, S.~He and E.~Y.~Yuan,
  ``Scattering of Massless Particles in Arbitrary Dimension,''
  arXiv:1307.2199 [hep-th].
  %%CITATION = ARXIV:1307.2199;%%
  %8 citations counted in INSPIRE as of 14 Nov 2013


%\cite{Dolan:2013isa}
\bibitem{Dolan:2013isa}
  L.~Dolan and P.~Goddard,
  ``Proof of the Formula of Cachazo, He and Yuan for Yang-Mills Tree Amplitudes in Arbitrary Dimension,''
  arXiv:1311.5200 [hep-th].
  %%CITATION = ARXIV:1311.5200;%%

%\cite{Britto:2005fq}
\bibitem{Britto:2005fq}
  R.~Britto, F.~Cachazo, B.~Feng and E.~Witten,
  ``Direct proof of tree-level recursion relation in Yang-Mills theory,''
  Phys.\ Rev.\ Lett.\  {\bf 94} (2005) 181602
  [hep-th/0501052].
  %%CITATION = HEP-TH/0501052;%%
  %540 citations counted in INSPIRE as of 20 Nov 2013

%\cite{Adamo:2013tsa}
\bibitem{Adamo:2013tsa}
  T.~Adamo, E.~Casali and D.~Skinner,
  ``Ambitwistor strings and the scattering equations at one loop,''
  arXiv:1312.3828 [hep-th].
  %%CITATION = ARXIV:1312.3828;%%
  %2 citations counted in INSPIRE as of 20 Jan 2014


%\cite{Berkovits:2013xba}
\bibitem{Berkovits:2013xba}
  N.~Berkovits,
  ``Infinite Tension Limit of the Pure Spinor Superstring,''
  arXiv:1311.4156 [hep-th].
  %%CITATION = ARXIV:1311.4156;%%

%\cite{Gross:1987kza}
\bibitem{Gross:1987kza}
  D.~J.~Gross and P.~F.~Mende,
  ``The High-Energy Behavior of String Scattering Amplitudes,''
  Phys.\ Lett.\ B {\bf 197} (1987) 129.
  %%CITATION = PHLTA,B197,129;%%
  %398 citations counted in INSPIRE as of 14 Nov 2013

%\cite{Gross:1987ar}
\bibitem{Gross:1987ar}
  D.~J.~Gross and P.~F.~Mende,
  ``String Theory Beyond the Planck Scale,''
  Nucl.\ Phys.\ B {\bf 303} (1988) 407.
  %%CITATION = NUPHA,B303,407;%%
  %633 citations counted in INSPIRE as of 14 Nov 2013


\bibitem{BZ-null}
%\cite{Bandos:1989nt}\bibitem{Bandos:1989nt}
  I.~A.~Bandos and A.~A.~Zheltukhin,
  ``Hamiltonian mechanics and absence of critical dimensions for null membranes,''
  Sov.\ J.\ Nucl.\ Phys.\  {\bf 50} (1989) 556
   [Yad.\ Fiz.\  {\bf 50} (1989) 893].
  %%CITATION = SJNCA,50,556;%%
  %13 citations counted in INSPIRE as of 14 Nov 2013

\bibitem{BZ-nulS}
%\cite{Bandos:1990mw} \bibitem{Bandos:1990mw}
 I.~A.~Bandos and A.~A.~Zheltukhin,
  ``Twistors, harmonics, and zero super-p-branes,''
  JETP Lett.\  {\bf 51} (1990) 618;
   %[Pisma Zh.\ Eksp.\ Teor.\ Fiz.\  {\bf 51} (1990) 547].
  %%CITATION = JTPLA,51,618;%%
  %16 citations counted in INSPIRE as of 14 Nov 2013
%\cite{Bandos:1991jx} \bibitem{Bandos:1991jx}  I.~A.~Bandos and A.~A.~Zheltukhin,
  ``Quantum theory of closed null supermembranes in four-dimensional space,''
  JETP Lett.\  {\bf 53} (1991) 5;
   %[Pisma Zh.\ Eksp.\ Teor.\ Fiz.\  {\bf 53} (1991) 7].
  %%CITATION = JTPLA,53,5;%%
  %12 citations counted in INSPIRE as of 14 Nov 2013
%\cite{Bandos:1991my}\bibitem{Bandos:1991my}  I.~A.~Bandos and A.~A.~Zheltukhin,
  ``Null super p-brane: Hamiltonian dynamics and quantization,''
  Phys.\ Lett.\ B {\bf 261} (1991) 245;
  %%CITATION = PHLTA,B261,245;%%
  %29 citations counted in INSPIRE as of 14 Nov 2013
%\cite{Bandos:1993ma}\bibitem{Bandos:1993ma}  I.~A.~Bandos and A.~A.~Zheltukhin,
  ``Null super p-branes quantum theory in four-dimensional space-time,''
  Fortsch.\ Phys.\  {\bf 41} (1993) 619.
  %%CITATION = FPYKA,41,619;%%
  %62 citations counted in INSPIRE as of 14 Nov 2013


%\cite{Schild:1976vq}
\bibitem{Schild:1976vq}
  A.~Schild,
  ``Classical Null Strings,''
  Phys.\ Rev.\ D {\bf 16}, 1722 (1977).
  %%CITATION = PHRVA,D16,1722;%%
  %236 citations counted in INSPIRE as of 07 Mar 2014
%\cite{Karlhede:1986wb}
\bibitem{Karlhede:1986wb}
  A.~Karlhede and U.~Lindstrom,
  ``The Classical Bosonic String in the Zero Tension Limit,''
  Class.\ Quant.\ Grav.\  {\bf 3} (1986) L73.
  %%CITATION = CQGRD,3,L73;%%
  %82 citations counted in INSPIRE as of 07 Mar 2014



%\cite{Zheltukhin:1997wj}
\bibitem{Zheltukhin:1997wj}
  A.~A.~Zheltukhin,
  ``A Hamiltonian of null strings: An invariant action of null (super)membranes,''
  Sov.\ J.\ Nucl.\ Phys.\  {\bf 48} (1988) 375
   [Yad.\ Fiz.\  {\bf 48} (1988) 587].
  %%CITATION = SJNCA,48,375;%%
  %53 citations counted in INSPIRE as of 07 Mar 2014


%\cite{Lindstrom:1990qb}
\bibitem{Lindstrom:1990qb}
  U.~Lindstrom, B.~Sundborg and G.~Theodoridis,
  ``The Zero tension limit of the superstring,''
  Phys.\ Lett.\ B {\bf 253} (1991) 319.
  %%CITATION = PHLTA,B253,319;%%
  %47 citations counted in INSPIRE as of 07 Mar 2014


%\cite{Bozhilov:1999mh}
\bibitem{Bozhilov:1999mh}
%\cite{Bozhilov:1998tn} \bibitem{Bozhilov:1998tn}
  P.~Bozhilov,
  ``N=1, D = 10 tensionless superbranes 1,''
  Phys.\ Lett.\ B {\bf 440} (1998) 35
  [hep-th/9806134].
  %%CITATION = HEP-TH/9806134;%%
  %4 citations counted in INSPIRE as of 01 Apr 2014
%\cite{Bozhilov:1999qe} \bibitem{Bozhilov:1999qe}
 %P.~Bozhilov,
  ``N=1, D = 10 tensionless superbranes II,''
  Phys.\ Lett.\ B {\bf 454} (1999) 27
  [hep-th/9901153].
  %%CITATION = HEP-TH/9901153;%%
  %4 citations counted in INSPIRE as of 01 Apr 2014
 %P.~Bozhilov,
  ``Null branes in curved backgrounds,''
  Phys.\ Rev.\ D {\bf 60} (1999) 125011
  [hep-th/9904208].
  %%CITATION = HEP-TH/9904208;%%
  %8 citations counted in INSPIRE as of 01 Apr 2014



%\cite{Lindstrom:2003mg}
\bibitem{Lindstrom:2003mg}
  U.~Lindstrom and M.~Zabzine,
  ``Tensionless strings, WZW models at critical level and massless higher spin fields,''
  Phys.\ Lett.\ B {\bf 584} (2004) 178
  [hep-th/0305098].
  %%CITATION = HEP-TH/0305098;%%
  %55 citations counted in INSPIRE as of 14 Nov 2013



\bibitem{BZ-str0}
I. A. Bandos and A. A. Zheltukhin, ``Spinor Cartan Moving N Hedron, Lorentz Harmonic Formulations Of Superstrings, And Kappa Symmetry,''
  JETP Lett.\  {\bf 54}, 421-424 (1991);
  %[Pisma Zh.\ Eksp.\ Teor.\ Fiz.\  {\bf 54}, 421 (1991)];
  %%CITATION = ZFPRA,54,421;%%
\bibitem{BZ-str}
 I.~A.~Bandos and A.~A.~Zheltukhin
  {\it Green-Schwarz
       superstrings in spinor moving frame formalism},
  Phys.\ Lett.\  {\bf B288}, 77-83 (1992);
  %%CITATION = PHLTA,B288,77;%%
%\bibitem{BZ-strH}
% I.~A.~Bandos and A.~A.~Zheltukhin
{\it D = 10 superstring:
Lagrangian and Hamiltonian mechanics in twistor-like Lorentz harmonic formulation},
Phys.\ Part.\ Nucl.\ {\bf 25} (1994) 453-477 [Preprint IC-92-422, ICTP, Trieste, 1992,
81pp.]
% {EChAYa} {\bf 25} (1994) No 5, p.1065--1127]
 %%CITATION = PPNUE,25,453;%%



\bibitem{Sok}
E.~Sokatchev, {\it Light cone harmonic cuperspace and its
applications},
  Phys.\ Lett.\  {\bf B169}, 209-214 (1986);
  %%CITATION = PHLTA,B169,209;%%
  {\it Harmonic superparticle},
  Class.\ Quant.\ Grav.\  {\bf 4}, 237-246 (1987).
  %%CITATION = CQGRD,4,237;%%



\bibitem{niss} E.~R.~Nissimov and S.~J.~Pacheva,
  ``Manifestly Superpoincare Covariant Quantization Of The Green-Schwarz Superstring,''
  Phys.\ Lett.\ B {\bf 202} (1988) 325;
  %%CITATION = PHLTA,B202,325;%%
E.~Nissimov, S.~Pacheva and S.~Solomon,
  ``Covariant Canonical Quantization Of The Green-Schwarz Superstring,''
  Nucl.\ Phys.\ B {\bf 297} (1988) 349;
  %%CITATION = NUPHA,B297,349;%%
``Off-shell Superspace D = 10 Super Yang-Mills From Covariantly Quantized Green-Schwarz Superstring,''
  Nucl.\ Phys.\ B {\bf 317} (1989) 344.
  %%CITATION = NUPHA,B317,344;%%



\bibitem{K+R88}
  R.~Kallosh and M.~Rakhmanov,
  ``Covariant Quantization Of The Green-schwarz Superstring,''
  Phys.\ Lett.\ B {\bf 209} (1988) 233;
  %%CITATION = PHLTA,B209,233;%%
  ``Consistency Of Covariant Quantization Of Gs String,''
  Phys.\ Lett.\ B {\bf 214} (1988) 549.
  %%CITATION = PHLTA,B214,549;%%

\bibitem{Ghsds}
  A.~S.~Galperin, P.~S.~Howe and K.~S.~Stelle, {\it The superparticle and the Lorentz
  group},  Nucl.\ Phys.\  {\bf B368}, 248-280 (1992)
  [hep-th/9201020];
  %%CITATION = NUPHA,B368,248;%%
\\ F.~Delduc, A.~Galperin and E.~Sokatchev,
  {\it Lorentz harmonic (super)fields and (super)particles},
  Nucl.\ Phys.\  B {\bf 368}, 143-171 (1992).
  %%CITATION = NUPHA,B368,143;%%


\bibitem{GHT93}
 A.~S.~Galperin, P.~S.~Howe and P.~K.~Townsend,
  {\it Twistor transform for superfields},
  Nucl.\ Phys.\  {\bf B402}, 531 (1993).
  %%CITATION = NUPHA,B402,531;%%


\bibitem{Francia:2002pt}
  D.~Francia and A.~Sagnotti,
  ``On the geometry of higher spin gauge fields,''
  Class.\ Quant.\ Grav.\  {\bf 20} (2003) S473
  [hep-th/0212185].
  %%CITATION = HEP-TH/0212185;%%
  %142 citations counted in INSPIRE as of 14 Nov 2013

%\cite{Sagnotti:2003qa}
\bibitem{Sagnotti:2003qa}
  A.~Sagnotti and M.~Tsulaia,
  ``On higher spins and the tensionless limit of string theory,''
  Nucl.\ Phys.\ B {\bf 682} (2004) 83
  [hep-th/0311257].
  %%CITATION = HEP-TH/0311257;%%
  %152 citations counted in INSPIRE as of 21 Nov 2013

%\cite{Bonelli:2003kh}
\bibitem{Bonelli:2003kh}
  G.~Bonelli,
  ``On the tensionless limit of bosonic strings, infinite symmetries and higher spins,''
  Nucl.\ Phys.\ B {\bf 669} (2003) 159
  [hep-th/0305155].
  %%CITATION = HEP-TH/0305155;%%
  %77 citations counted in INSPIRE as of 14 Nov 2013



%\cite{Abe:2004ep}
\bibitem{Abe:2004ep}
  Y.~Abe, V.~P.~Nair and M.~-I.~Park,
  ``Multigluon amplitudes, N =4 constraints and the WZW model,''
  Phys.\ Rev.\ D {\bf 71} (2005) 025002
  [hep-th/0408191].
  %%CITATION = HEP-TH/0408191;%%
  %36 citations counted in INSPIRE as of 14 Nov 2013



 %\cite{Boels:2006ir}
\bibitem{Boels:2006ir}
  R.~Boels, L.~J.~Mason and D.~Skinner,
  ``Supersymmetric Gauge Theories in Twistor Space,''
  JHEP {\bf 0702} (2007) 014
  [hep-th/0604040].
  %%CITATION = HEP-TH/0604040;%%
  %77 citations counted in INSPIRE as of 14 Nov 2013

 %\cite{Adamo:2013cra}
\bibitem{Adamo:2013cra}
  T.~Adamo,
  ``Twistor actions for gauge theory and gravity,'' PhD Thesis, Univ. of Oxford,
  arXiv:1308.2820 [hep-th].
  %%CITATION = ARXIV:1308.2820;%%
  %1 citations counted in INSPIRE as of 14 Nov 2013






%\cite{Bandos:2006af}
\bibitem{Bandos:2006af}
  I.~A.~Bandos, J.~A.~de Azcarraga and C.~Miquel-Espanya,
  ``Superspace formulations of the (super)twistor string,''
  JHEP {\bf 0607} (2006) 005
  [hep-th/0604037].
  %%CITATION = HEP-TH/0604037;%%
  %20 citations counted in INSPIRE as of 14 Nov 2013




%\cite{Siegel:2004dj}
\bibitem{Siegel:2004dj}
  W.~Siegel,
  ``Untwisting the twistor superstring,''
  hep-th/0404255.
  %%CITATION = HEP-TH/0404255;%%
  %61 citations counted in INSPIRE as of 14 Nov 2013



%\cite{Witten:2003nn}
\bibitem{Witten:2003nn}
  E.~Witten,
  ``Perturbative gauge theory as a string theory in twistor space,''
  Commun.\ Math.\ Phys.\  {\bf 252} (2004) 189
  [hep-th/0312171].
  %%CITATION = HEP-TH/0312171;%%
  %706 citations counted in INSPIRE as of 14 Nov 2013


\bibitem{Berkovits:2004hg}
  N.~Berkovits,
  ``An Alternative string theory in twistor space for N=4 superYang-Mills,''
  Phys.\ Rev.\ Lett.\  {\bf 93} (2004) 011601
  [hep-th/0402045].
  %%CITATION = HEP-TH/0402045;%%
  %148 citations counted in INSPIRE as of 14 Nov 2013

%\cite{Engelund:2014sqa}
\bibitem{Engelund:2014sqa}
  O.~T.~Engelund and R.~Roiban,
  ``A twistor string for the ABJ(M) theory,''
  arXiv:1401.6242 [hep-th].
  %%CITATION = ARXIV:1401.6242;%%


\bibitem{B90}
%\cite{Bandos:1990ji} \bibitem{Bandos:1990ji}
  I.~A.~Bandos,
  ``Superparticle in Lorentz harmonic superspace,''
  Sov.\ J.\ Nucl.\ Phys.\  {\bf 51} (1990) 906
   % [Yad.\ Fiz.\  {\bf 51} (1990) 1429 in Russian].
  %%CITATION = SJNCA,51,906;%%
  %85 citations counted in INSPIRE as of 02 Apr 2014


\bibitem{IB+AN=95}
%\cite{Bandos:1996ju}\bibitem{Bandos:1996ju}
  I.~A.~Bandos and A.~Y.~Nurmagambetov,
  ``Generalized action principle and extrinsic geometry for N=1 superparticle,''
  Class.\ Quant.\ Grav.\  {\bf 14} (1997) 1597
  [hep-th/9610098].
  %%CITATION = HEP-TH/9610098;%%

%\cite{Bandos:1998wj}
\bibitem{Bandos:1998wj}
  I.~A.~Bandos and J.~Lukierski,
  ``New superparticle models outside the HLS supersymmetry scheme,''
  Lect.\ Notes Phys.\  {\bf 539} (2000) 195
  [hep-th/9812074].
  %%CITATION = HEP-TH/9812074;%%
  %23 citations counted in INSPIRE as of 26 Mar 2014


%\cite{Bandos:2006nr,IB07:M0}
\bibitem{Bandos:2006nr}
  I.~A.~Bandos, J.~A.~de Azcarraga and D.~P.~Sorokin,
  ``On D=11 supertwistors, superparticle quantization and a hidden SO(16) symmetry of supergravity,''  in: "Quantum, Super and Twistors, Proc. XXII Max Born Symposium, Wroclaw (Poland) 2006", Eds: J. Kowalski-Glikman and Ludwik Turko,  Wroclaw University Press  2008, pp. 25-32 [hep-th/0612252].
  %%CITATION = HEP-TH/0612252;%%
  %8 citations counted in INSPIRE as of 14 Feb 2014

\bibitem{IB07:M0}
 I.~A.~Bandos,
  ``Spinor moving frame, M0-brane covariant BRST quantization and intrinsic complexity
  of the pure spinor approach,''
  Phys.\ Lett.\ B {\bf 659} (2008) 388
  [arXiv:0707.2336 [hep-th]];
  %%CITATION = ARXIV:0707.2336;%%
 I.A. Bandos,
``D=11 massless superparticle covariant quantization, pure spinor BRST charge
  and hidden symmetries,''
  Nucl. Phys. {\bf B796}, 360 (2008).
  %[arXiv:0710.4342 [hep-th]].
  %%CITATION = NUPHA,B796,360;%%




%\cite{Siegel:1983hh}\bibitem{Siegel:1983hh}
 \bibitem{kappaS}
  W.~Siegel,
  ``Hidden Local Supersymmetry in the Supersymmetric Particle Action,''
  Phys.\ Lett.\ B {\bf 128}, 397 (1983).
  %%CITATION = PHLTA,B128,397;%%

\bibitem{kappaAL}
%\cite{de Azcarraga:1982dw}\bibitem{de Azcarraga:1982dw}
  J.~A.~de Azcarraga and J.~Lukierski,
 ``Supersymmetric Particles with Internal Symmetries and Central Charges,''
  Phys.\ Lett.\ B {\bf 113}, 170 (1982);
    %%CITATION = PHLTA,B113,170;%%
 Phys.\ Rev.\ D {\bf 28}, 1337 (1983).
  %%CITATION = PHRVA,D28,1337;%%


%\cite{Bergshoeff:1996tu}
\bibitem{Bergshoeff:1996tu}
  E.~Bergshoeff and P.~K.~Townsend,
  ``Super D-branes,''
  Nucl.\ Phys.\ B {\bf 490} (1997) 145
  [hep-th/9611173].
  %%CITATION = HEP-TH/9611173;%%
  %403 citations counted in INSPIRE as of 20 Nov 2013


%\cite{Boulanger:2011dd}
\bibitem{Boulanger:2011dd}
  N.~Boulanger and P.~Sundell,
  ``An action principle for Vasiliev's four-dimensional higher-spin gravity,''
  J.\ Phys.\ A {\bf 44}, 495402 (2011)
  [arXiv:1102.2219 [hep-th]].
  %%CITATION = ARXIV:1102.2219;%%
  %35 citations counted in INSPIRE as of 11 Feb 2014

%\cite{Boulanger:2012bj}
\bibitem{Boulanger:2012bj}
  N.~Boulanger, N.~Colombo and P.~Sundell,
  ``A minimal BV action for Vasiliev's four-dimensional higher spin gravity,''
  JHEP {\bf 1210}, 043 (2012)
  [arXiv:1205.3339 [hep-th]].
  %%CITATION = ARXIV:1205.3339;%%
  %7 citations counted in INSPIRE as of 11 Feb 2014

\bibitem{Vasiliev:1988sa}
%\cite{Vasiliev:1988xc}\bibitem{Vasiliev:1988xc}
  M.~A.~Vasiliev,
  ``Equations of Motion of Interacting Massless Fields of All Spins as a Free Differential Algebra,''
  Phys.\ Lett.\ B {\bf 209}, 491 (1988);
  %%CITATION = PHLTA,B209,491;%%
  %72 citations counted in INSPIRE as of 11 Feb 2014
%\cite{Vasiliev:1988sa}\bibitem{Vasiliev:1988sa}  M.~A.~Vasiliev,
  ``Consistent Equations for Interacting Massless Fields of All Spins in the First Order in Curvatures,''
  Annals Phys.\  {\bf 190} (1989) 59;
  %%CITATION = APNYA,190,59;%%
  %160 citations counted in INSPIRE as of 11 Feb 2014
%\cite{Vasiliev:1990en} \bibitem{Vasiliev:1990en}  M.~A.~Vasiliev,
  ``Consistent equation for interacting gauge fields of all spins in (3+1)-dimensions,''
  Phys.\ Lett.\ B {\bf 243}, 378 (1990).
  %%CITATION = PHLTA,B243,378;%%
  %281 citations counted in INSPIRE as of 11 Feb 2014

%\cite{Fronsdal:1978rb}
\bibitem{Fronsdal:1978rb}
  C.~Fronsdal,
  ``Massless Fields with Integer Spin,''
  Phys.\ Rev.\ D {\bf 18}, 3624 (1978).
  %%CITATION = PHRVA,D18,3624;%%
  %457 citations counted in INSPIRE as of 03 Apr 2014



%\cite{Ferber:1977qx}
\bibitem{Ferber:1977qx}
  A.~Ferber,
  ``Supertwistors and Conformal Supersymmetry,''
  Nucl.\ Phys.\ B {\bf 132}, 55--64 (1978).
  %%CITATION = NUPHA,B132,55;%%

  %\cite{Shirafuji:1983zd}
  \bibitem{Shirafuji:1983zd}
  T.~Shirafuji,
  ``Lagrangian Mechanics Of Massless Particles With Spin,''
  Prog.\ Theor.\ Phys.\  {\bf 70}, 18--35 (1983).
  %%CITATION = PTPKA,70,18;%%
%%%%DOI: 10.1143/PTP.70.18





%\cite{Newman:1961qr}
\bibitem{Newman:1961qr}
  E.~Newman and R.~Penrose,
  %``An Approach to gravitational radiation by a method of spin coefficients,''
  J.\ Math.\ Phys.\  {\bf 3} (1962) 566.
  %%CITATION = JMAPA,3,566;%%
  %775 citations counted in INSPIRE as of 07 Mar 2014

  %\cite{Penrose:1987uia}
\bibitem{Penrose:1987uia}
  R.~Penrose and W.~Rindler,
  ``Spinors And Space-time. 1. Two Spinor Calculus And Relativistic Fields,''
  Cambridge, Uk: Univ. Pr. ( 1984) 458 pp. ( Cambridge Monographs On Mathematical Physics);
  %7 citations counted in INSPIRE as of 07 Mar 2014
%\cite{Penrose:1986ca} \bibitem{Penrose:1986ca}  R.~Penrose and W.~Rindler,
  ``Spinors And Space-time. Vol. 2: Spinor And Twistor Methods In Space-time Geometry,''
  Cambridge, Uk: Univ. Pr. ( 1986) 501p.
  %14 citations counted in INSPIRE as of 07 Mar 2014

\bibitem{Uvarov}
%\cite{Uvarov:2006ed}\bibitem{Uvarov:2006ed}
  D.~V.~Uvarov,
  ``(Super)twistors and (super)strings,''
  Class.\ Quant.\ Grav.\  {\bf 23} (2006) 2711
  [hep-th/0601149];
  %%CITATION = HEP-TH/0601149;%%
  %10 citations counted in INSPIRE as of 07 Mar 2014
%\cite{Uvarov:2007vs}\bibitem{Uvarov:2007vs}   D.~V.~Uvarov,
  ``Supertwistor formulation for higher dimensional superstrings,''
  Class.\ Quant.\ Grav.\  {\bf 24} (2007) 5383
  [hep-th/0703051 [HEP-TH]];
  %%CITATION = HEP-TH/0703051;%%
  %5 citations counted in INSPIRE as of 07 Mar 2014
%\cite{Uvarov:2008bx} \bibitem{Uvarov:2008bx} D.~V.~Uvarov,
  ``Canonical description of D=10 superstring formulated in supertwistor space,''
  J.\ Phys.\ A {\bf 42} (2009) 115204
  [arXiv:0804.0908 [hep-th]].
  %%CITATION = ARXIV:0804.0908;%%
  %1 citations counted in INSPIRE as of 07 Mar 2014

\bibitem{Fedoruk}
%\cite{Fedoruk:2006de}\bibitem{Fedoruk:2006de}
  S.~Fedoruk and J.~Lukierski,
  ``Twistorial versus space-time formulations: Unification of various string models,''
  Phys.\ Rev.\ D {\bf 75} (2007) 026004
  [hep-th/0606245];
  %%CITATION = HEP-TH/0606245;%%
  %11 citations counted in INSPIRE as of 07 Mar 2014
%\cite{Fedoruk:2008qr} \bibitem{Fedoruk:2008qr}
 % S.~Fedoruk and J.~Lukierski,
  ``Purely twistorial string with canonical twistor field quantization,''
  Phys.\ Rev.\ D {\bf 79} (2009) 066006
  [arXiv:0811.3353 [hep-th]].
  %%CITATION = ARXIV:0811.3353;%%

%\cite{Bandos:2013uoa}
\bibitem{Bandos:2013uoa}
  I.~A.~Bandos and C.~Meliveo,
  ``Covariant action and equations of motion for the eleven dimensional multiple M0-brane system,''
  Phys.\ Rev.\ D {\bf 87} (2013) 12,  126011
  [arXiv:1304.0382 [hep-th]].
  %%CITATION = ARXIV:1304.0382;%%
  %1 citations counted in INSPIRE as of 28 Mar 2014



%\cite{Berkovits:2004jj}
\bibitem{Berkovits:2004jj}
  N.~Berkovits and E.~Witten,
  ``Conformal supergravity in twistor-string theory,''
  JHEP {\bf 0408} (2004) 009
  [hep-th/0406051].
  %%CITATION = HEP-TH/0406051;%%
  %146 citations counted in INSPIRE as of 04 Apr 2014

%\cite{AbouZeid:2006wu}
\bibitem{AbouZeid:2006wu}
  M.~Abou-Zeid, C.~M.~Hull and L.~J.~Mason,
  ``Einstein Supergravity and New Twistor String Theories,''
  Commun.\ Math.\ Phys.\  {\bf 282} (2008) 519
  [hep-th/0606272].
  %%CITATION = HEP-TH/0606272;%%
  %39 citations counted in INSPIRE as of 28 Mar 2014






%\cite{vanHolten:1982mx}
\bibitem{vanHolten:1982mx}
  J.~W.~van Holten and A.~Van Proeyen,
  ``N=1 Supersymmetry Algebras in D=2, D=3, D=4 MOD-8,''
  J.\ Phys.\ A {\bf 15} (1982) 3763.
  %%CITATION = JPHGB,A15,3763;%%
  %165 citations counted in INSPIRE as of 02 Apr 2014

%\cite{Curtright:1987zc}
\bibitem{Curtright:1987zc}
  T.~Curtright,
  ``Are There Any Superstrings in Eleven-dimensions?,''
  Phys.\ Rev.\ Lett.\  {\bf 60} (1988) 393
   [Erratum-ibid.\  {\bf 60} (1988) 1990].
  %%CITATION = PRLTA,60,393;%%
  %36 citations counted in INSPIRE as of 02 Apr 2014

%\cite{Bandos:1998vz}
\bibitem{Bandos:1998vz}
  I.~A.~Bandos and J.~Lukierski,
  ``Tensorial central charges and new superparticle models with fundamental spinor coordinates,''
  Mod.\ Phys.\ Lett.\ A {\bf 14} (1999) 1257
  [hep-th/9811022].
  %%CITATION = HEP-TH/9811022;%%
  %106 citations counted in INSPIRE as of 02 Apr 2014



%\cite{Chryssomalakos:1999xd}
\bibitem{Chryssomalakos:1999xd}
  C.~Chryssomalakos, J.~A.~de Azcarraga, J.~M.~Izquierdo and J.~C.~Perez Bueno,
  ``The Geometry of branes and extended superspaces,''
  Nucl.\ Phys.\ B {\bf 567} (2000) 293
  [hep-th/9904137].
  %%CITATION = HEP-TH/9904137;%%
  %63 citations counted in INSPIRE as of 02 Apr 2014

%\cite{Bandos:2001pu}
\bibitem{Bandos:2001pu}
  I.~A.~Bandos, J.~A.~de Azcarraga, J.~M.~Izquierdo and J.~Lukierski,
  ``BPS states in M theory and twistorial constituents,''
  Phys.\ Rev.\ Lett.\  {\bf 86} (2001) 4451
  [hep-th/0101113].
  %%CITATION = HEP-TH/0101113;%%
  %67 citations counted in INSPIRE as of 02 Apr 2014

%\cite{Bandos:2003ng}
\bibitem{Bandos:2003ng}
  I.~A.~Bandos, J.~A.~de Azcarraga, M.~Picon and O.~Varela,
  ``Supersymmetric string model with 30 kappa symmetries and extended supersapce and 30|32 BPS states,''
  Phys.\ Rev.\ D {\bf 69} (2004) 085007
  [hep-th/0307106].
  %%CITATION = HEP-TH/0307106;%%
  %30 citations counted in INSPIRE as of 02 Apr 2014



%\cite{D'Auria:1982nx}
\bibitem{D'Auria:1982nx}
  R.~D'Auria and P.~Fre,
  ``Geometric Supergravity in d = 11 and Its Hidden Supergroup,''
  Nucl.\ Phys.\ B {\bf 201} (1982) 101
   [Erratum-ibid.\ B {\bf 206} (1982) 496].
  %%CITATION = NUPHA,B201,101;%%
  %193 citations counted in INSPIRE as of 02 Apr 2014

%\cite{Bandos:2004xw}
\bibitem{Bandos:2004xw}
  I.~A.~Bandos, J.~A.~de Azcarraga, J.~M.~Izquierdo, M.~Picon and O.~Varela,
  ``On the underlying gauge group structure of D=11 supergravity,''
  Phys.\ Lett.\ B {\bf 596} (2004) 145
  [hep-th/0406020].
  %%CITATION = HEP-TH/0406020;%%
  %24 citations counted in INSPIRE as of 02 Apr 2014


%\cite{Bandos:2004ym}
\bibitem{Bandos:2004ym}
  I.~A.~Bandos, J.~A.~de Azcarraga, M.~Picon and O.~Varela,
  ``On the formulation of D = 11 supergravity and the composite nature of its three-form gauge field,''
  Annals Phys.\  {\bf 317} (2005) 238
  [hep-th/0409100].
  %%CITATION = HEP-TH/0409100;%%
  %24 citations counted in INSPIRE as of 02 Apr 2014

%\cite{Bandos:2001pu}
\bibitem{Bandos:2001pu}
  I.~A.~Bandos, J.~A.~de Azcarraga, J.~M.~Izquierdo and J.~Lukierski,
  ``BPS states in M theory and twistorial constituents,''
  Phys.\ Rev.\ Lett.\  {\bf 86} (2001) 4451
  [hep-th/0101113].
  %%CITATION = HEP-TH/0101113;%%
  %67 citations counted in INSPIRE as of 02 Apr 2014

  %\cite{Bandos:2003us}
\bibitem{Bandos:2003us}
  I.~A.~Bandos, J.~A.~de Azcarraga, J.~M.~Izquierdo, M.~Picon and O.~Varela,
  ``On BPS preons, generalized holonomies and D = 11 supergravities,''
  Phys.\ Rev.\ D {\bf 69} (2004) 105010
  [hep-th/0312266].
  %%CITATION = HEP-TH/0312266;%%
  %46 citations counted in INSPIRE as of 02 Apr 2014


%\cite{Bjerrum-Bohr:2014qwa}
\bibitem{Bjerrum-Bohr:2014qwa}
  N.~E.~JBjerrum-Bohr, P.~H.~Damgaard, P.~Tourkine and P.~Vanhove,
  ``Scattering Equations and String Theory Amplitudes,''
  arXiv:1403.4553 [hep-th].
  %%CITATION = ARXIV:1403.4553;%%

%\cite{Geyer:2014fka}
\bibitem{Geyer:2014fka}
  Y.~Geyer, A.~E.~Lipstein and L.~J.~Mason,
  ``Ambitwistor strings in 4-dimensions,''
  arXiv:1404.6219 [hep-th].
  %%CITATION = ARXIV:1404.6219;%%

\end{thebibliography}
\end{document}